\documentclass[lettersize,journal]{IEEEtran}
\usepackage{amsmath,amsfonts,amssymb}
\usepackage{algorithm}
\usepackage{array}
\usepackage[caption=false,font=normalsize,labelfont=sf,textfont=sf]{subfig}
\usepackage{textcomp}
\usepackage{stfloats}
\usepackage{url}
\usepackage{verbatim}
\usepackage{graphicx}
\usepackage{cite,color,multirow}

\usepackage{hyperref}

\usepackage{algpseudocode}

\usepackage{booktabs}

\def\ve#1{{\mathchoice{\mbox{\boldmath$\displaystyle #1$}}%
{\mbox{\boldmath$\textstyle #1$}}%
{\mbox{\boldmath$\scriptstyle #1$}}%
{\mbox{\boldmath$\scriptscriptstyle #1$}}}}
\def\j{\mathrm{j}}

\def\duallsas{\lambda_{\mathrm{LSAS}}}
\def\ledi{\lambda_{\mathrm{EDI}}}
\def\lrds{\lambda_{\mathrm{RDS}}}
\def\lrdst{\tilde\lambda_{\mathrm{RDS}}}
\def\x{\mathrm{x}}
\def\y{\mathrm{y}}
\def\spm{\mathrm{spm}}
\def\xpm{\mathrm{xpm}}

\begin{document}

\title{Probabilistic Amplitude Shaping and Nonlinearity Tolerance: Analysis and Sequence Selection Method}

\author{Mohammad Taha Askari, Lutz Lampe, ~\IEEEmembership{Senior Member,~IEEE}, Jeebak Mitra
\thanks{M.T. Askari and L. Lampe are with the Department
of Electrical and Computer Engineering, The University of British Columbia, Vancouver,
BC V6T 1Z4, Canada, e-mail: (mohammadtaha@ece.ubc.ca, lampe@ece.ubc.ca).}
\thanks{J. Mitra was with Huawei Technologies Canada, Ottawa, He is now with Dell Technologies Canada, e-mail:  jeebak.mitra@dell.com.}
}

\markboth{}%
{Shell \MakeLowercase{\textit{et al.}}: A Sample Article Using IEEEtran.cls for IEEE Journals}


\maketitle

\begin{abstract}
Probabilistic amplitude shaping (PAS) is a practical means to  achieve a shaping  gain in optical fiber communication. However, PAS and shaping in general also affect the signal-dependent generation of nonlinear interference. This provides an opportunity for nonlinearity mitigation through PAS, which is also referred to as a nonlinear shaping gain. In this paper, we introduce a linear lowpass filter model that relates transmitted symbol-energy sequences and nonlinear distortion experienced in an optical fiber channel. Based on this model, we conduct a nonlinearity analysis of PAS with respect to shaping blocklength and mapping strategy. Our model explains results and relationships found in literature and can be used as a design tool for PAS with improved nonlinearity tolerance. We use the model to introduce a new metric for PAS with sequence selection. We perform simulations of selection-based PAS  with various amplitude shapers and mapping strategies to demonstrate the effectiveness of the new metric in different optical fiber system scenarios.
\end{abstract}


\section{Introduction}
\label{intro}
\IEEEPARstart{P}{robabilistic} constellation shaping is a well-established approach to improve over uniform signaling providing an ultimate shaping gain of 1.53~dB for the additive white Gaussian noise (AWGN) channel \cite{fischer:2005}. Among the various probabilistic shaping methods \cite{gultekin2020probabilistic}, PAS has emerged as an effective design for integrating shaping and forward error correction (FEC) coding \cite{bocherer2015bandwidth}. The PAS architecture consists of an inner FEC encoder concatenated with an outer amplitude shaper (AS). The AS maps uniformly distributed input bits to amplitudes and generates a non-uniform amplitude distribution. It operates on finite-length blocks of amplitudes at a time, which results in a rate loss of the achievable information rate (AIR) compared to the infinite blocklength limit.

In addition to the conventional \textsl{linear} shaping gain, shaping also has the potential to exhibit a \textsl{nonlinear} gain compared to uniform signaling for the optical fiber channel \cite{dar2014shaping}. 
Several previous works, e.g., \cite{hansen2020optimization,cho2022kurtosis,pan2016probabilistic,fehenberger2016probabilistic}, attributed the nonlinear shaping gain to a change in the moments and in particular the kurtosis of the amplitudes distribution. For PAS, it has furthermore been demonstrated in \cite{fehenberger2019analysis,fehenberger2020mitigating,fehenberger2020impact} that the effective signal-to-noise ratio (SNR) depends on the blocklength of the AS. As the AS blocklength increases, the nonlinear shaping gain decreases, while the linear shaping gain increases. Since this trend is not observed if a symbol interleaver is applied to shaped sequences, it can be concluded that the temporal properties of the shaped sequences play a role in the severity of nonlinear interference (NLI). 

Besides the shaping blocklength, the choice of mapping of amplitudes from one shaping block to the components of the transmitted signal affects the nonlinearity tolerance of PAS transmission. In particular, it has been reported that mapping shaping blocks across in-phase and quadrature-phase components in single-polarization transmission and also across ${\mathrm{x}}$- and ${\mathrm{y}}$-polarization in dual-polarization transmission can increase the effective SNR \cite{skvortcov2021huffman}. This effect can be explained by noting that the optical channel acts as a lowpass filter on symbol-energy sequences causing NLI, and that finite blocklength PAS produces a spectral dip at frequency zero for such sequences \cite{peng2020transmission}. The filter point of view also facilitates understanding the effect of AS together with other system parameters such as baud rate on NLI \cite{peng2021baud}. The significance of the properties of symbol-energy sequences for the NLI in PAS transmission is further highlighted by the so-called energy dispersion index (EDI) introduced in \cite{wu2021temporal}. The EDI is a measure for the variance of the windowed energy of the shaped signal, where the window length is infinite in the case of the exponentially weighted EDI (EEDI) \cite{wu2021exponentially}. For PAS implemented with constant composition distribution matching (CCDM) \cite{schulte2015constant}, both EEDI and EDI have been demonstrated to be good predictors for effective SNR.

The observations of the interplay between PAS and NLI suggest a modification of the traditional AS design that includes the nonlinear shaping gain as an objective. Two types of such modified designs have emerged in the literature. The first pairs a conventional AS with a selection module, in which the latter chooses from a candidate set of shaped amplitude sequences generated by the AS. The selection is aided by a criterion that accounts for the nonlinearity of the optical channel, such as the EDI in the list-encoding CCDM (L-CCDM) of \cite{wu2022list} or the effective SNR obtained from channel emulation using the split-step Fourier method (SSFM) in {  \cite{secondini2022new}}. The generation of candidate sets can be integrated into the operation of a conventional AS by inserting additional flipping bits into each information block that is input to the AS. The second approach for nonlinearity tolerant shaping directly modifies the sequence generation in the conventional AS. For example, the trellis of the enumerative sphere shaping (ESS) algorithm \cite{amari2019introducing} can be adjusted to obtain sequences with improved nonlinearity tolerance. This has been done based on the per-sequence kurtosis in K-ESS \cite{gultekin2021kurtosis} and considering sequence-energy variations in band-trellis ESS (B-ESS) \cite{gultekin2022mitigating}. The bisection-based implementation of CCDM (BS-CCDM) presented in \cite{fu2021parallel} exploits the redundancy in CCDM to remove unfavorable CCDM sequences and thus mitigate fiber nonlinearity. 

In this paper, we revisit the interplay of PAS and fiber nonlinearity using the first-order perturbation based model for NLI. First, we develop an effective \textsl{linear channel model} for the conversion of symbol-energy sequences into NLI. Our model extends the lowpass filter model from  \cite{peng2020transmission} and accounts for self-phase modulation (SPM) and cross-phase modulation (XPM) as well as intra-polarization and inter-polarization effects. Using the frequency domain representation of the effective linear channel, we provide insights into the nonlinearity tolerance of finite-length PAS as a function of system parameters such as link length and baud rate. This analysis permits us to explain several phenomenological observations reported in previous works, such as the optimization of AS blocklength and mapping strategy to maximize the effective SNR. As a second contribution, we employ our channel model to introduce the lowpass-filtered symbol-amplitude sequence (LSAS) metric for nonlinearity tolerant PAS using sequence selection.\footnote{We note that a special case of the LSAS metric was originally introduced in our conference paper \cite{taha:2022}.} We identify its relation to the EDI, which is interesting in itself as it validates the EDI as a metric for nonlinearity tolerance of shaped sequences. It furthermore highlights that different from the EDI, the derivation of the LSAS metric does not rely on simplifying assumptions for quantifying NLI caused by symbol-energy sequences, and it generalizes systematically to XPM and inter-polarization NLI. The effectiveness of the new metric for PAS with sequence selection is demonstrated through simulations of single-span and multi-span fiber links for single-polarization and dual-polarization setups, which show that the LSAS metric outperforms the EDI metric in terms of effective SNR, AIR, and error rate. Finally, we integrate the different strategies for improved nonlinearity tolerance with PAS. That is, for the first time, we combine (i) multidimensional mapping and (ii) nonlinearity tolerant K-ESS with LSAS-based sequence selection. Our numerical results demonstrate (i) that multidimensional mapping is effective in reducing the search space and thus complexity for PAS with sequence selection and (ii) that the combination of K-ESS and sequence selection yields notable benefits in terms of, e.g.,  effective SNR for the nonlinear fiber channel, compared to applying them individually. 

\begin{figure*}[!t]
\centering
\includegraphics[width=7in]{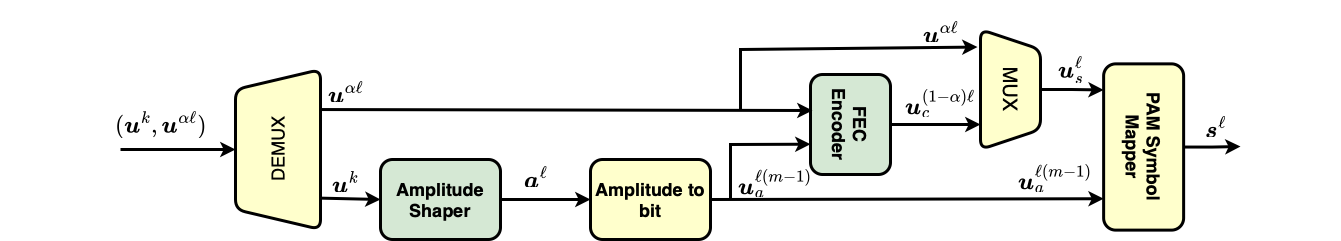}
\caption{Block diagram for conventional PAS. Superscripts denote the sequence length. $\ve{u}$, $\ve{a}$, $\ve{s}$ denote a bit, PAM amplitude, and PAM symbol sequences, respectively.}
\label{fig:PAS_block_diagram}
\end{figure*}

The remainder of this paper is organized as follows. In Section~\ref{system_model}, we briefly explain the system model for PAS, symbol mapping, and sequence selection. In Section~\ref{energy_model}, we derive the linear channel model between symbol-energy sequences and NLI using perturbation theory, and we discuss the effect of fiber link and system parameters on the channel properties. We then introduce the LSAS metric in Section~\ref{lsas}, we show how the EDI can be derived as an approximation of LSAS,  {  and we discuss the computational complexity associated with sequence selection using EDI and LSAS}. In Section~\ref{results}, we present quantitative comparisons of different nonlinearity tolerant PAS methods. We highlight the benefits of the LSAS-metric and its combination with multidimensional mapping and different AS methods by simulating both short-haul and long-haul optical fiber transmission scenarios. {  In Section~\ref{sec:discussion}, we reflect on aspects of our work that merit further consideration and are thus suggested as future research. } Section~\ref{conclusion} concludes the paper.

\section{Probabilistic Amplitude Shaping,  Mapping, and Sequence Selection}
\label{system_model}
In this section, we briefly review conventional PAS and PAS with sequence selection. We also discuss the options for mapping symbols to the signal components in dual-polarized optical fiber transmission.

\subsection{Probabilistic Amplitude Shaping (PAS)}
\label{sec:pas}
Figure~\ref{fig:PAS_block_diagram} shows  the operation of PAS following the description in \cite{bocherer2015bandwidth}.
The AS maps a block $\ve{u}^k$ of $k$ independent and uniformly distributed information bits to a block $\ve{a}^\ell$ of $\ell$ amplitudes of a $2^{m}$-ary pulse-amplitude modulation (PAM) constellation. The amplitudes of the PAM symbols have a marginal distribution $P_{\mathrm a}$. The elements of $\ve{a}^\ell$ are amplitude-to-bit converted into the sequence $\ve{u}_{\mathrm{a}}^{\ell(m-1)}$ of  $\ell(m-1)$ amplitude bits. A sequence $\ve{u}_{\mathrm{s}}^{\ell}$ of $\ell$ sign bits is obtained from $\alpha\ell$ information bits $\ve{u}^{\alpha\ell}$, $0 \leq \alpha \leq 1$ , and $(1-\alpha)\ell$ bits $\ve{u}_{\mathrm{c}}^{(1-\alpha)\ell}$ generated using a systematic FEC encoder. Each $m$ bits consisting of $m-1$ amplitude bits from $\ve{u}_{\mathrm{a}}^{\ell(m-1)}$ and one sign bit from $\ve{u}_{\mathrm{s}}^{\ell}$ are mapped to a PAM symbol resulting in $\ell$ symbols $\ve{s}^{\ell}$. 

The shaping rate of the system is $R_{\mathrm{s}} = \frac{k}{\ell}$, and the overall data rate of this PAS scheme is  $R = R_{\mathrm{s}} + \alpha$. The rate loss is defined as 
\begin{equation}
\label{eq:rateloss}
R_{\mathrm{loss}} = H(P_{\mathrm{a}}) - R_{\mathrm{s}},
\end{equation}
where $H(P_{\mathrm{a}})$ is the entropy associated with the distribution $P_{\mathrm{a}}$.

\begin{figure}[!t]
\centering
\subfloat[]{\includegraphics[width=0.55\textwidth]{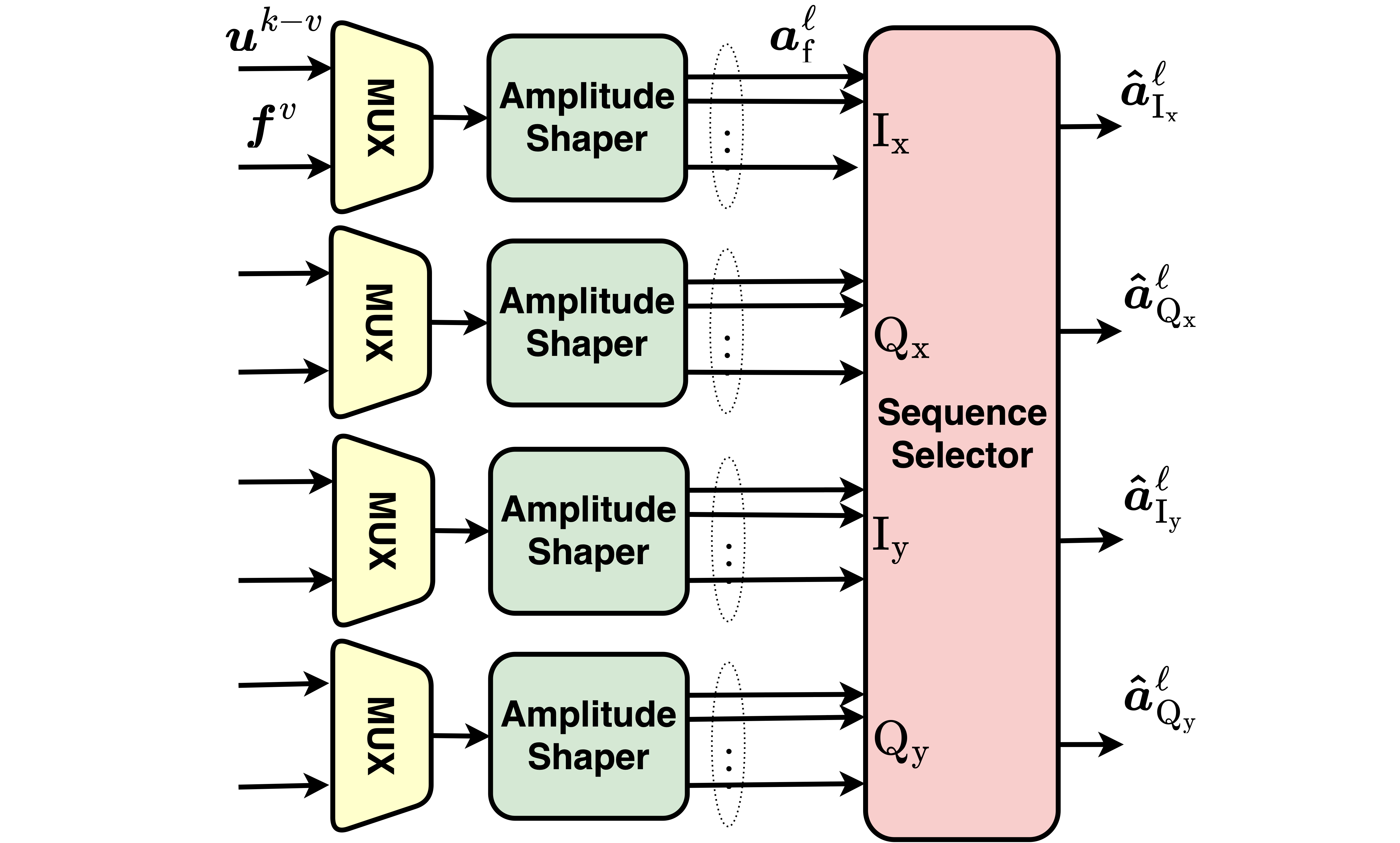}%
\label{1D}}
\newline
\subfloat[]{\includegraphics[width=0.6\textwidth]{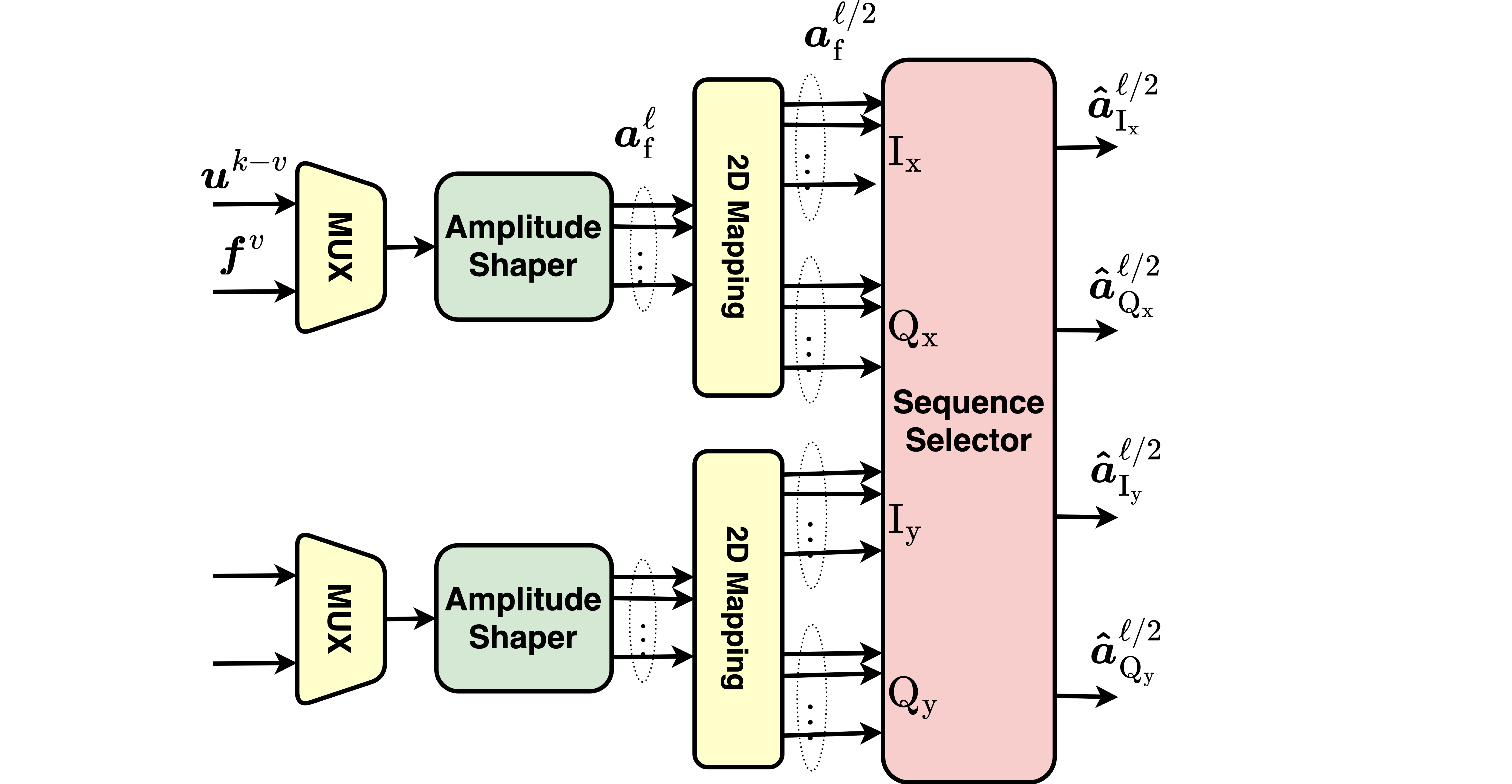}%
\label{2D}}
\newline
\subfloat[]{\includegraphics[width=0.55\textwidth]{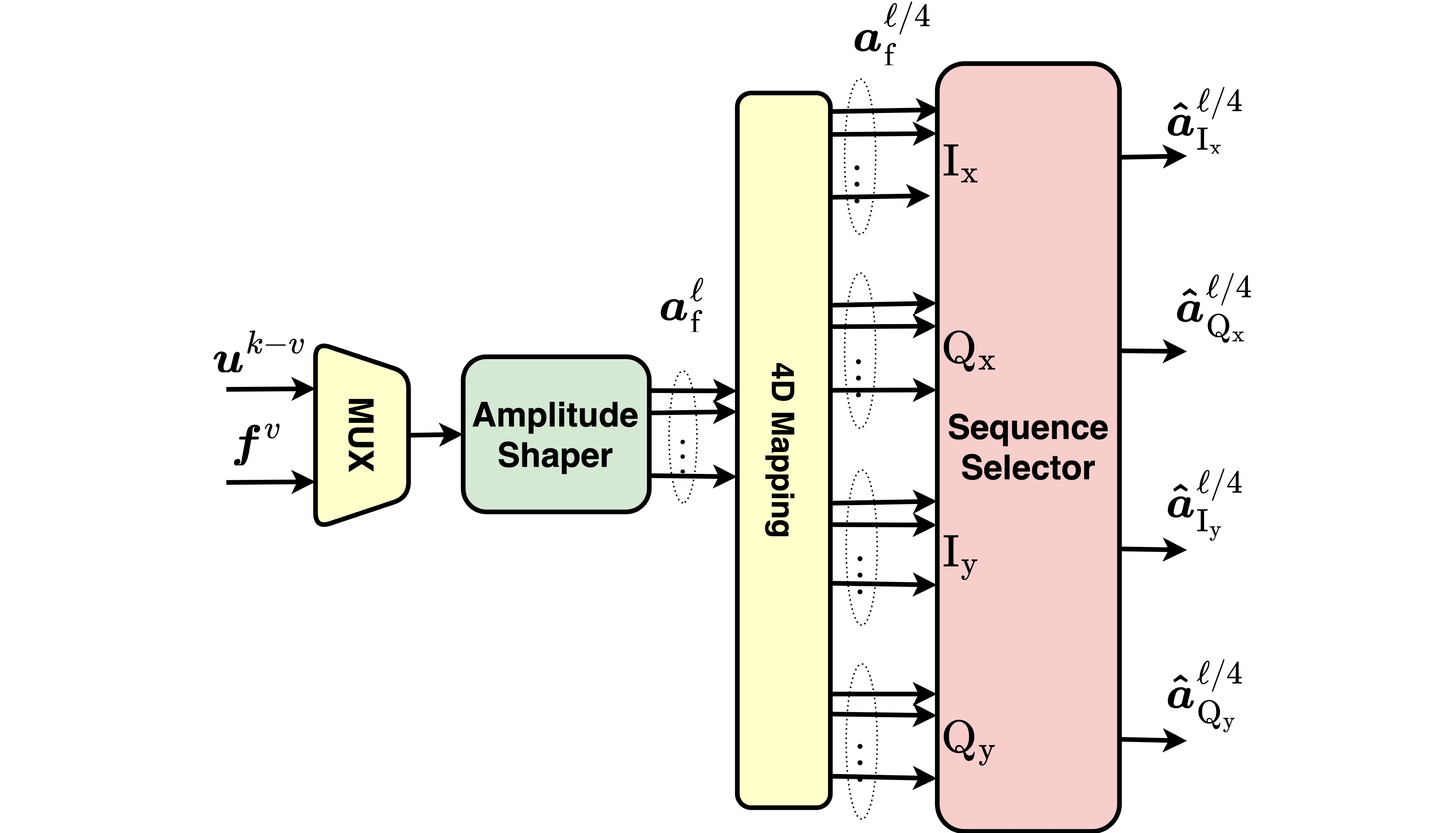}%
\label{4D}}
\caption{Block diagram for different mapping strategies: (a) 1D, (b) 2D, and (c) 4D mapping, followed by  sequence selection.}
\label{fig:sequence_selection_BL}
\end{figure}

\subsection{Mapping Strategy}
\label{s:mapping}
We consider dual-polarized optical fiber transmission using quadrature amplitude modulation (QAM). This means that PAM blocks $\ve{s}^\ell$ are mapped to four-dimensional (4D) symbol sequences, where each symbol has $\mathrm{I_x}$, $\mathrm{Q_x}$, $\mathrm{I_y}$, and $\mathrm{Q_y}$ components for in-phase and quadrature and ${\mathrm{x}}$- and ${\mathrm{y}}$-polarization dimensions, respectively. We differentiate three strategies for this mapping, which we refer to as 1D, 2D, and 4D mapping. For the case of 1D mapping, each shaping block $\ve{s}^\ell$ is mapped to a separate quadrature and polarization dimension. The 2D mapping strategy maps each shaping block across the two quadrature components but to a single polarization. Finally, each shaping block is mapped to all four 4D-symbol components in 4D mapping. The different mapping strategies are illustrated in Figure~\ref{fig:sequence_selection_BL} in terms of amplitude blocks $\ve{a}^\ell$, i.e., we map the shaped amplitude blocks to the signal components and apply the sign bits afterward, as this facilitates the discussion of sequence selection further below.  

The choice of mapping has an effect on nonlinearity tolerance with PAS for two reasons. First, the mapping strategy changes the time duration during which the total energy of one shaping block is transmitted. This can be seen in Figure~\ref{fig:sequence_selection_BL} comparing the lengths of the amplitude sequences at the output of the mappers. Thus, the mapping affects the effective shaping blocklength. Second, 2D and 4D mapping introduce dependencies between the different components of the 4D symbols. We will discuss the consequences of these interactions between the mapping and finite-length PAS in Section~\ref{s:filtermapping}. 

\subsection{Sequence Selection}
\label{s:sequenceselection}
When the AS of the PAS scheme produces several shaped amplitude blocks for representing the same block of information bits, the mapping step discussed in the previous section is followed by a sequence selector module (see Figure~\ref{fig:sequence_selection_BL}). 
{  
The generation of candidate shaping blocks is enabled by allocating the first $v$ out of $k$ bits as redundant bits $\ve{f}^v$, which are referred to as flipping bits \cite{wu2022list}. Flipping the first bits provides more variation to the set of generated candidates for common shaping methods such as CCDM and ESS. The remaining $k-v$ bits are considered as information bits $\ve{u}^{k-v}$. The candidate generation, which is adopted from \cite{wu2022list}, is independent of the selected amplitude shaper, i.e., the same process is applied to e.g.\ CCDM or ESS.
}
For each combination of flipping bits, a shaping block of $\ell$ PAM amplitude symbols $\ve{a}_{\mathrm{f}}^{\ell}$ is generated. The amplitudes are mapped to a block of $\ell/d$ 4D-amplitude symbols according to the $(d\in\{1,2,4\})$-dimensional mapping strategy as shown in Figure~\ref{fig:sequence_selection_BL}. 
The sequence selector computes a metric for each 4D candidate and selects the best candidate $\ve{\hat a}^{\ell/d}=[\ve{\hat a}^{\ell/d}_{\mathrm{I}_\mathrm{x}},\ldots,\ve{\hat a}_{\mathrm{Q}_\mathrm{y}}^{\ell/d}]$. We note that the size of the candidate set depends on the mapping strategy { for a fixed shaping blocklength $\ell$}. Since $4/d$ ASs contribute to the generation of 4D-amplitude symbols, the selection module computes $2^{\frac{4}{d}v}\times d$ metrics for producing $\ell$ 4D symbols. Hence, computational complexity decreases with the dimension $d$ of the mapping. After selection, amplitude-to-bit conversion and FEC operations are performed in the same way as for the conventional PAS scheme shown in Figure~\ref{fig:PAS_block_diagram}.

When introducing flipping bits, the rate loss expression \eqref{eq:rateloss} generalizes to
\begin{equation}
\label{eq:rateloss_sequence_selection}
R_{\mathrm{loss},v} = H(P_{\mathrm{a}}) - R_{\mathrm{s},v} = H(P_{\mathrm{a}}) - \frac{k-v}{\ell}.
\end{equation} 
In order to have a fixed effective rate $R=\frac{k-v}{\ell}+\alpha$,  $k-v$ should be remained fixed. Hence, as $v$ increases, $k$ also increases, resulting in lower linear shaping gain for given $\ell$. Therefore, sequence selection  is a mechanism to trade off linear  with nonlinear shaping gain.

\section{Linear Channel Approximation for Nonlinear Distortion}
\label{energy_model}
In this section, we derive the approximation of the nonlinear distortion as the output of a linear channel that is excited by the symbol-energy sequences as its input. We discuss the properties of the linear filter representing the channel as a function of fiber-link parameters. We then use this channel model to explain the effects of finite-length PAS and multidimensional mapping on the nonlinear tolerance of transmission. 

\subsection{Linear Channel Model Formulation}
\label{s:linearchannel}
We consider the optical fiber channel after compensation of linear distortions at the receiver and focus on nonlinear distortion from signal-to-signal interactions.
We assume dual-polarization transmission in a wavelength division multiplexing (WDM) system with a set of channels ${\cal C}$. The transmit symbol in polarization $p\in{\cal P}=\{\mathrm{x,y}\}$ and channel $c\in{\cal C}$ at discrete time $n\in\mathbb{Z}$ is denoted by $s_{p}^{(c)}(n)$, and the symbol energy is 
\begin{equation}
\label{eq:energy_def}
E_p^{(c)}(n)\stackrel{\Delta}{=}|s_{p}^{(c)}(n)|^2. 
\end{equation}
Then, using the first-order perturbation based distortion model, as briefly reviewed in the Appendix, and only retaining symbol-energy terms, we can approximate the corresponding received sample as
\begin{equation} 
\label{eq:energy_model_a}
\begin{split}
r^{(c)}_{p}(n) &\approx s^{(c)}_{p}(n) \bigg[1 \\
& \hspace*{-5mm}+ \j \gamma \underbrace{\bigg( \sum_{p'\in{\cal P}}\sum_{c'\in{\cal C} } \sum_{m\in\mathbb{Z}} E_{p'}^{(c')}(n+m)h^{(c,c')}_{p,p'}(m)\bigg)}_{\stackrel{\Delta}{=}D_p^{(c)}(n)} \bigg],
\end{split}
\end{equation}
where $\gamma$ is  the fiber nonlinearity parameter, and $h^{(c,c')}_{p,p'}(m)\in\mathbb{R}_{\ge 0}$ are coefficients determined by the fiber channel. We observe that the distortion term can be written as ($\ast$ denotes convolution) 
\begin{equation} 
\label{eq:dist}
D^{(c)}_{p}(n) = \sum_{p'\in{\cal P}}\sum_{c'\in{\cal C} } \left(E_{p'}^{(c')}\ast h^{(c,c')}_{p,p'}\right)(n),
\end{equation}
i.e., the summation of the linear convolution between the filters $h^{(c,c')}_{p,p'}$ and the symbol-energy sequences $E_{p'}^{(c')}$ for all WDM channels and both polarizations. The filters $h^{(c,c')}_{p,p'}$ represent intra- ($p=p')$ and inter-polarization ($p\neq p')$ interferences of SPM ($c=c'$) and XPM ($c\neq c'$), respectively.
We note that the linear channel model \eqref{eq:dist} is not meant to be sufficiently accurate for perturbation-based nonlinearity compensation. However, it is suitable for explaining the interplay between shaping and multidimensional mapping and NLI, and for guiding the design of nonlinearity tolerant PAS. 

It is insightful to develop \eqref{eq:energy_model_a} and \eqref{eq:dist} further, which also allows us to connect to other related work. For this, we introduce the deterministic mean distortion term
\begin{equation}
\bar{D}_p^{(c)} = \sum_{p'\in{\cal P}}\sum_{c'\in{\cal C} } \bar{E}_{p'}^{(c')} \sum_{m\in\mathbb{Z}}h^{(c,c')}_{p,p'}(m)
\end{equation}
and the distortion variation term
\begin{equation}   
\label{e:deltadist}
\Delta D_{p,p'}^{(c,c')}(n)= \left(\left[E_{p'}^{(c')}-\bar{E}_{p'}^{(c')}\right]\ast h^{(c,c')}_{p,p'}\right)(n),
\end{equation}
where  ($\mathbb{E}$ denotes statistical expectation)
{ 
\begin{equation}
\label{eq:expected_energy_def}
\bar{E}_p^{(c)}=\mathbb{E}\left[E_{p}^{(c)}(n)\right]
\end{equation}
}is the mean symbol energy. 
Then, the total distortion \eqref{eq:dist} decomposes as 
\begin{equation}   
\label{e:deltatwithdeltadist}
D^{(c)}_{p}(n) = \sum_{p'\in{\cal P}}\sum_{c'\in{\cal C} }\Delta D_{p,p'}^{(c,c')}(n)+\bar{D}_{p}^{(c)}\;.
\end{equation}
Applying \eqref{e:deltatwithdeltadist} and the first order Taylor series approximation of the exponential function to \eqref{eq:energy_model_a}, we obtain the approximation 
\begin{equation}
r^{(c)}_{p}(n) \approx s^{(c}_{p}(n) \exp\bigg[\j \gamma \bigg( \sum_{p'\in{\cal P}}
\sum_{c'\in{\cal C} } \Delta D_{p,p'}^{(c,c')}(n) + \bar{D}_p^{(c)}
\bigg) \bigg].
\label{eq:energy_model_b}
\end{equation}
The NLI is now approximated as phase noise, which suggests that a carrier phase recovery (CPR) may mitigate the distortion. Since the effectiveness of a CPR  depends on the dynamic of the distortion term, the temporal structure of the symbol-energy and thus the shaped amplitude sequence are important. Simulation results in \cite{civelli2020interplay, borujeny2022constant} suggest that a powerful CPR can be similarly effective in mitigating NLI as finite-length PAS, at least in the high SNR regime \cite{civelli2020interplay}. But even a mean-phase recovery would compensate for the deterministic term $\bar{D}_p^{(c)}$, and therefore we will exclude it in the design of a selection metric in Section~\ref{lsas}.

Furthermore, the relationship for the filter coefficients $h^{(c,c')}_{p,p}(n)$ provided in \eqref{eq:filter_coeff} in the Appendix motivates the approximation 
\begin{equation} \label{eq:filter_approximation}
    h^{(c,c')}_{p,p' \neq p}(n) \approx \frac{1}{2} h^{(c,c')}_{p,p}(n) \stackrel{\Delta}{=} h^{(c,c')}_{p}(n),\quad \forall c,c'\in{\cal C}.
\end{equation}
This leads to the simplified approximation
\begin{equation} \label{eq:delta_e_simp}
    D_p^{(c)}(n) \approx \sum_{c'\in{\cal C}} \left( \left[2E_p^{(c')}+E_{p'\neq p}^{(c')}\right]\ast h_p^{(c,c')}\right)(n),
\end{equation}
for \eqref{eq:dist}, where the nonlinear distortion is approximated as linear filtering of the sum of symbol-energy sequences for the two signal polarizations. The simplified formulation \eqref{eq:delta_e_simp} is equivalent to \cite[Eqs.~(3) and (4)]{peng2020transmission}.

\subsection{Filter Properties}
We now evaluate the expression for the filters $h_{p,p'}^{(c,c')}$ for selected optical fiber links. If not specified otherwise, the link parameters are as given in Table~\ref{tab:setup1_sim_param}.

\begin{figure*}[!t]
\centering
\subfloat[]{\includegraphics[width=0.5\textwidth]{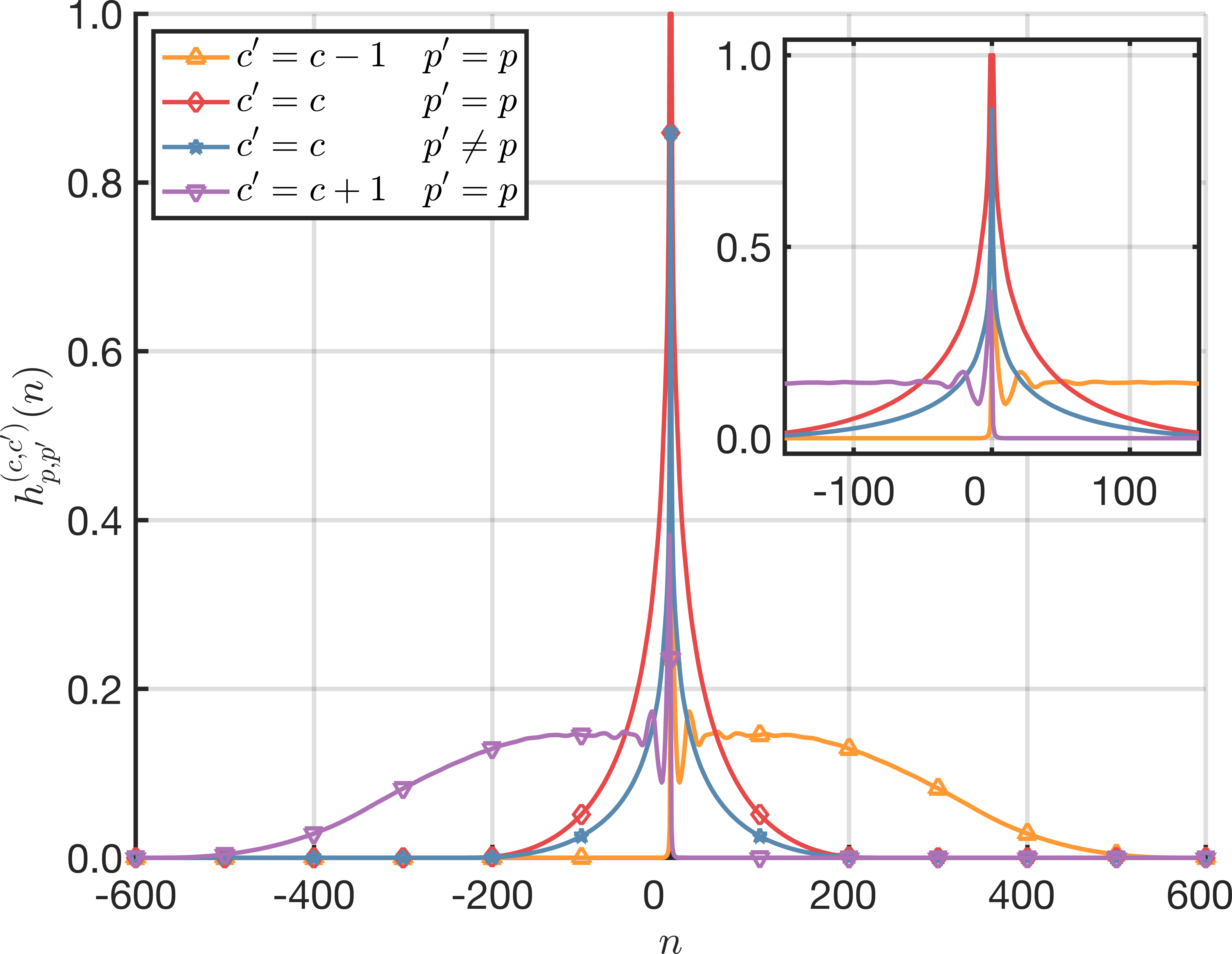}%
\label{fig:h_vs_n_a}}
\hfil
\subfloat[]{\includegraphics[width=0.5\textwidth]{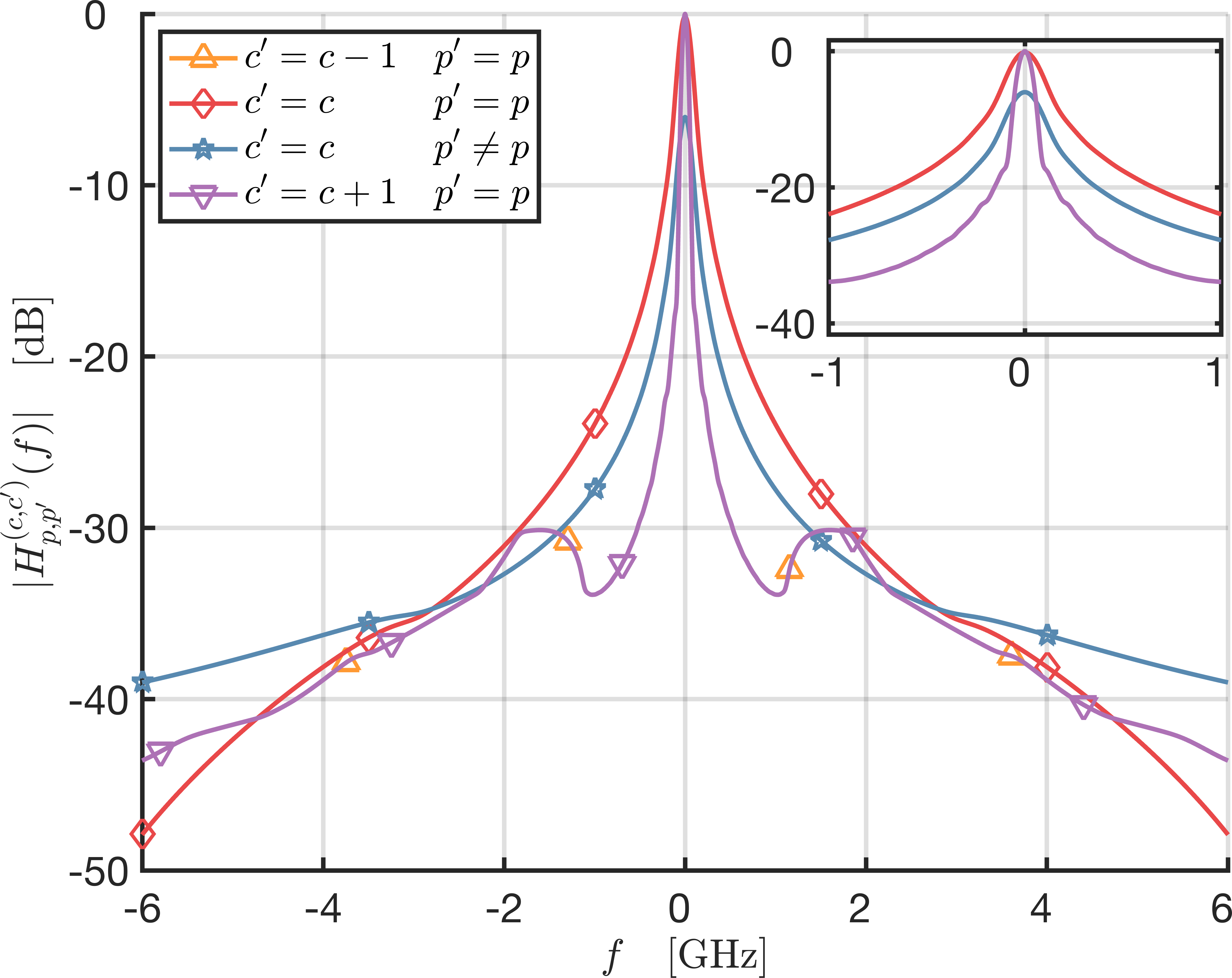}%
\label{fig:h_vs_f_b}}
\caption{Normalized filters for a fiber link with length $1600$~km, $32$~GBd, and $50$~GHz channel spacing. SPM filter (intra- and inter-polarization) and XPM (intra polarization) filter from two adjacent channels. (a) Time domain, maximum coefficient normalized to 1. (b) Frequency domain (magnitude), maximum normalized to $0$~dB at $f=0$. The insets show an interval around the origin.}
\label{fig:h_vs_n_f}
\end{figure*}

Figures~\ref{fig:h_vs_n_a} and~\ref{fig:h_vs_f_b} show the normalized filter coefficients and their normalized magnitude Fourier transforms, respectively, for SPM and for XPM from two adjacent channels for a $1600$~km link, i.e., 20 spans of length 80~km. We observe that all filters have a lowpass characteristic and that the XPM filters ($c\neq c'$) are more frequency-selective than their SPM counterparts ($c=c'$). This suggests that the modulation of the temporal characteristics of amplitude sequences via shaping will be more effective in suppressing NLI from XPM than from SPM. We also observe from the two SPM filters in Figure~\ref{fig:h_vs_n_a} that the intra-polarization coefficients are two times higher than the inter-polarization coefficients except for $n=0$ (see Appendix). The two XPM filters for  $c'=c+1$ and $c'=c-1$ in Figure~\ref{fig:h_vs_n_a} are time-reversed functions and therefore their magnitude frequency responses in Figure~\ref{fig:h_vs_f_b} overlap completely.

\begin{figure}[!t]
\centering
\includegraphics[width=0.47\textwidth]{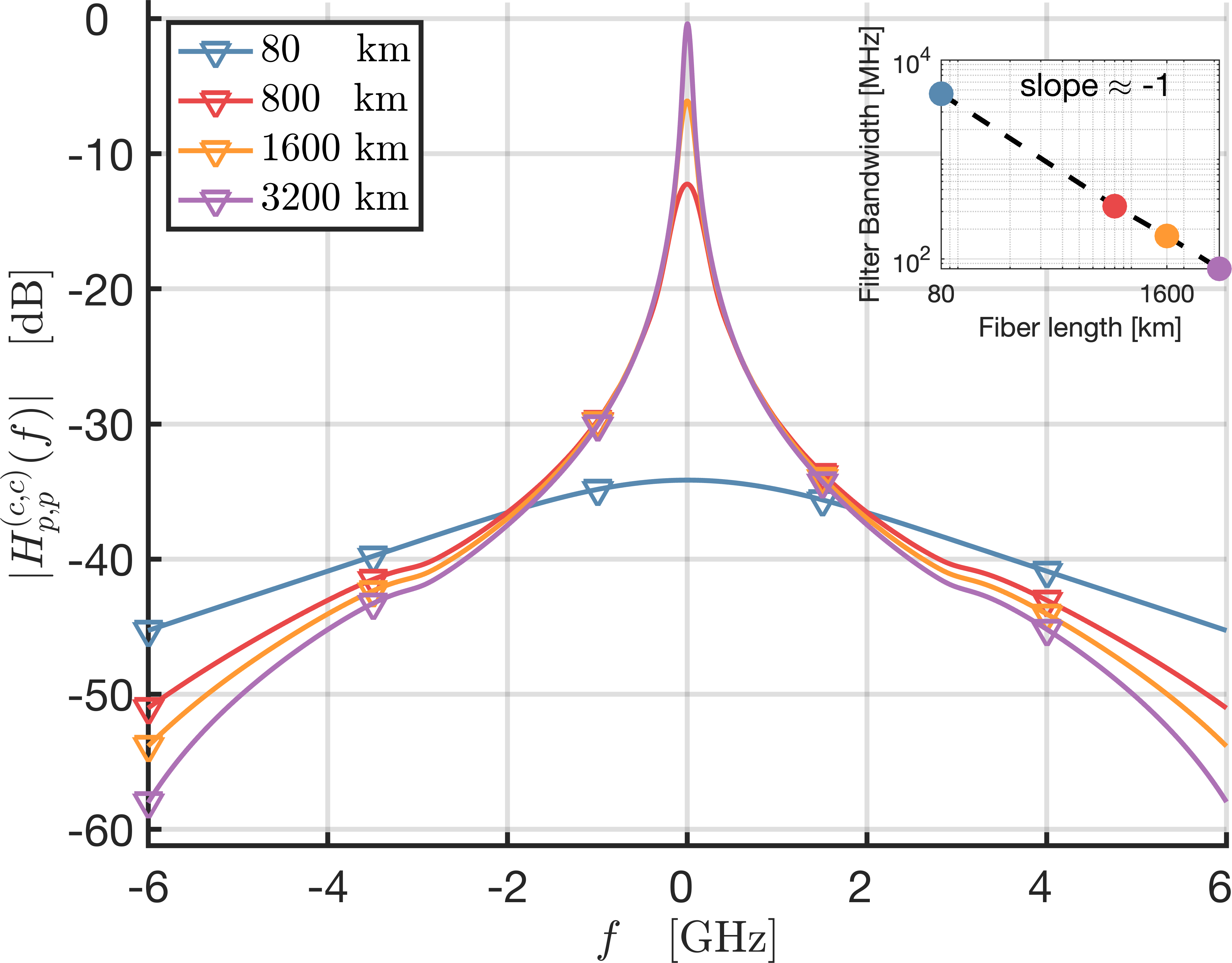}
\caption{Magnitude frequency response for intra-polarization SPM filter for a transmission with $32$~GBd and $50$~GHz channel spacing over fibers with various link lengths. Normalized to a maximum of $0$~dB at $f=0$. Inset: $3$~dB filter bandwidth vs.\ fiber link length. The corresponding bandwidth of each filter is shown with the same color code.}
\label{fig:h_vs_f_different_lengths}
\end{figure}

\begin{figure}[!t]
\centering
\includegraphics[width=0.5\textwidth]{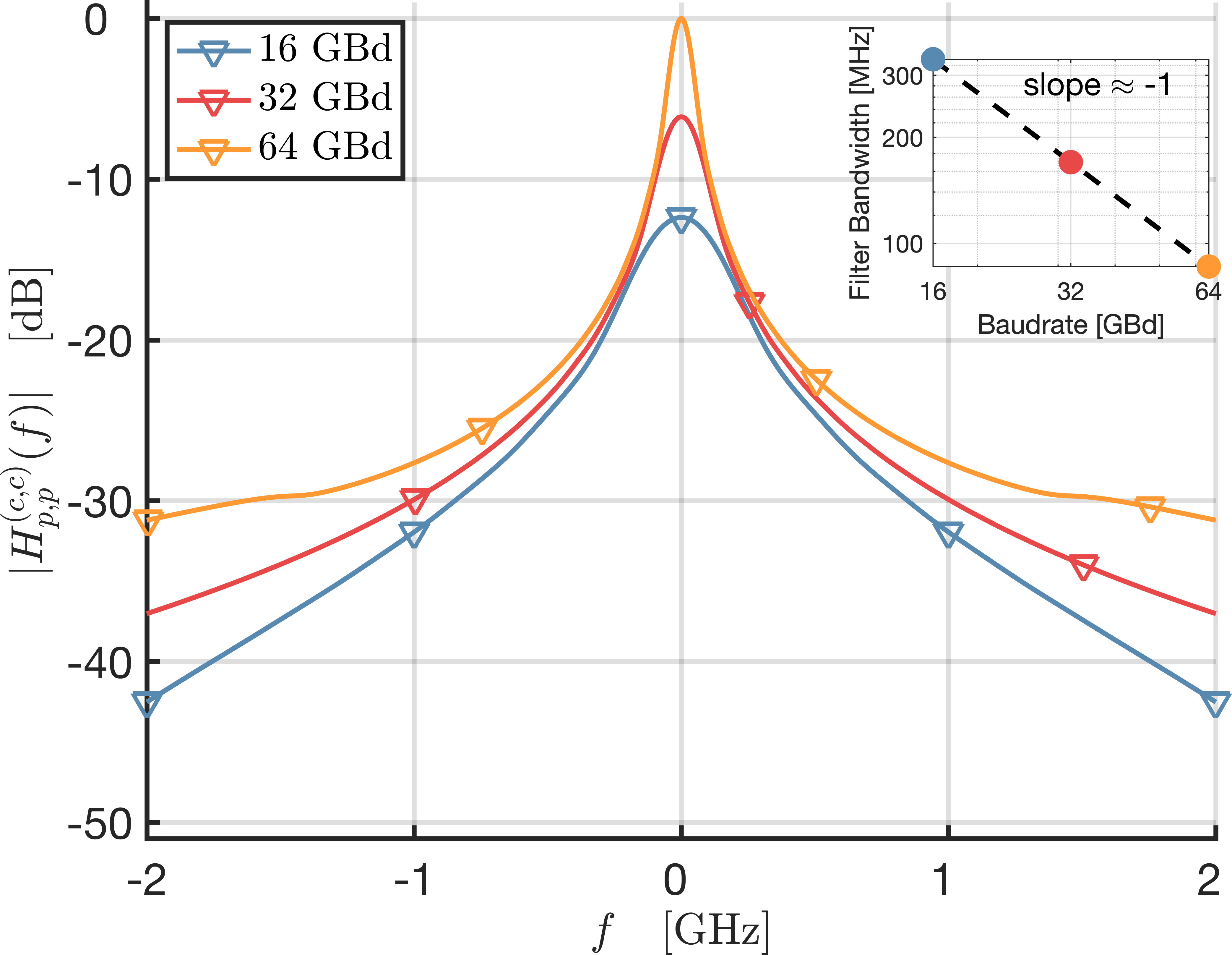}
\caption{Magnitude frequency response for intra-polarization SPM filter for a transmission over a $1600$~km fiber link with three different baud rates. Normalized to a maximum of $0$~dB at $f=0$. Inset: $3$~dB filter bandwidth vs.\ baud rate. The corresponding bandwidth of each filter is shown with the same color code.}
\label{fig:h_vs_f_different_baudrate}
\end{figure}

Figure~\ref{fig:h_vs_f_different_lengths} shows the normalized magnitude frequency response for the intra-polarization SPM filters for fiber links with $1$, $10$, $20$, and $40$ spans of $80$~km. 
As the link length increases, the channel memory increases, and as a result, the lowpass filters become more selective. For this, the inset in  Figure~\ref{fig:h_vs_f_different_lengths} shows the  $3$~dB filter bandwidth versus the fiber link length in a log-log plot. We observe that the trend is linear with an approximate slope of $-1$. This is consistent with channel memory being linearly proportional to link length, e.g.,\ \cite[Eq.~(2)]{wu2021temporal}. Similar observations apply for XPM filters, which we have not shown for brevity. Furthermore, we note the higher magnitude of filter frequency responses at zero for longer links, which indicates a more substantial overall nonlinear distortion.

Next, we consider a fiber link with a fixed length of $1600$~km and different baud rates of $16$~GBd, $32$~GBd, and $64$~GBd. Figure~\ref{fig:h_vs_f_different_baudrate} shows the normalized magnitude frequency response for the intra-polarization SPM filter for these cases. A larger baud rate and thus signal bandwidth results in a more distinct lowpass characteristic in the frequency domain. The inset figure shows the $3$~dB filter bandwidth versus transmission baud rate in a log-log plot. We observe that the bandwidth decreases as the baud rate increases with a linear slope of about $-1$. Again, this is  consistent with channel memory, measured in time duration, being linearly proportional to baud rate, e.g.,\ \cite[Eq.~(2)]{wu2021temporal}.
We also expect more significant nonlinear SPM effects with increasing baud rate, which is confirmed by the larger frequency response value at zero.  Based on the observations in Figures~\ref{fig:h_vs_f_different_lengths} and~\ref{fig:h_vs_f_different_baudrate}, we suggest the relationship $\mathrm{BW} \propto \frac{1}{L\cdot B}$ for the $3$~dB bandwidth of the SPM intra-polarization filter, fiber link length $L$, and baud rate $B$.

\subsection{Interplay Between Filter and PAS Blocklength}
\label{spectral_dip}
The linear channel model is an effective tool to provide insights into the role of PAS in the generation of NLI. In particular, from the channel model  \eqref{eq:dist} and the lowpass filter characteristics illustrated in the previous section, we conclude that symbol-energy sequences with weaker low-frequency components would result in less NLI. Such an argument has already been made in \cite{peng2020transmission} using the simplified approximation \eqref{eq:delta_e_simp} for comparing multidimensional mappings in conjunction with finite-length PAS. 

\begin{figure}[!t]
\centering
\includegraphics[width=0.5\textwidth]{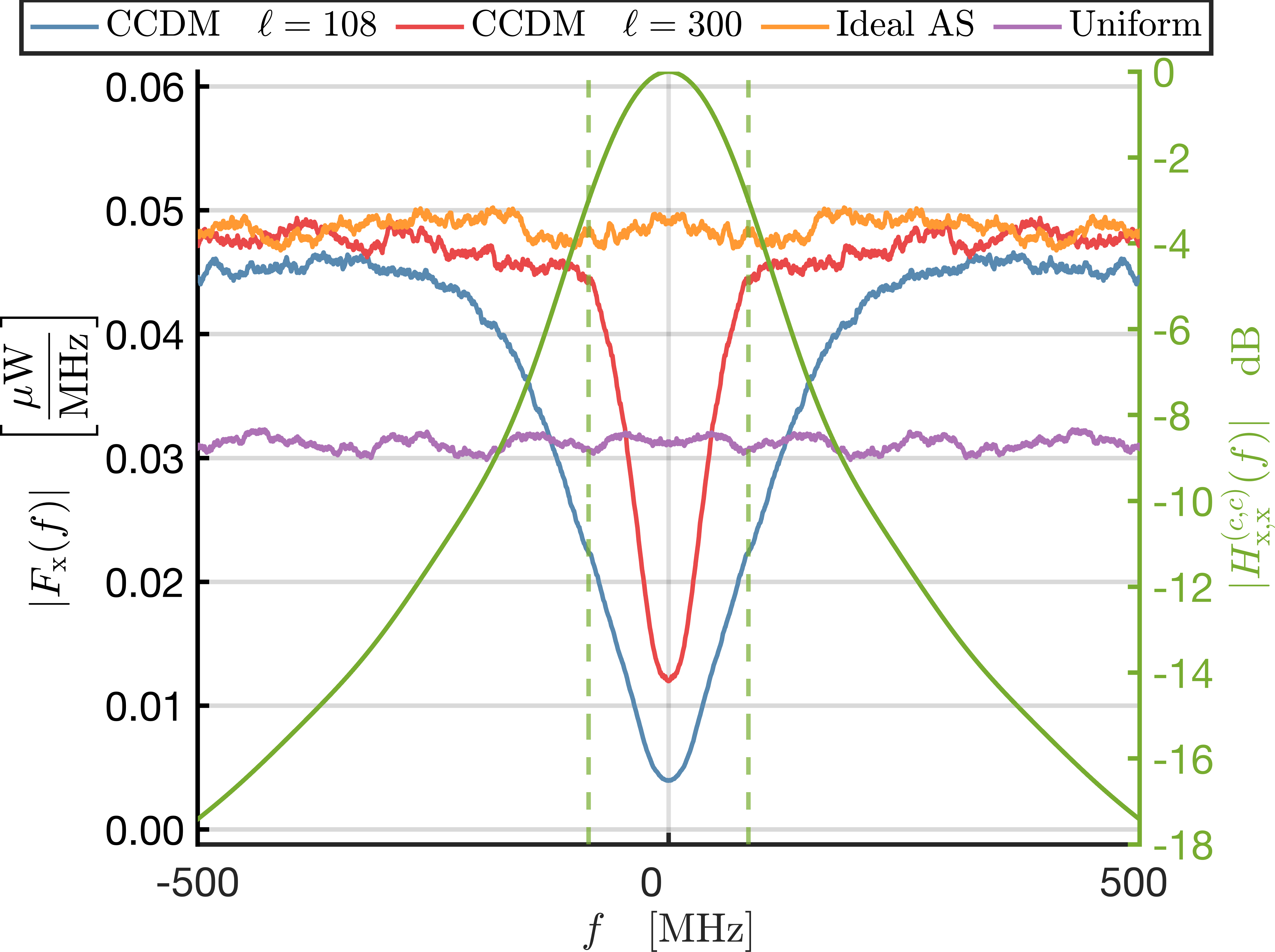}
\caption{Left: Frequency domain representation of $\mathrm{x}$-polarization symbol-energy sequences generated by CCDM with blocklength $108$ and $300$, and by uniform signaling and ideal AS. The launch power is $-6.5$~dBm per polarization. Right: Normalized magnitude frequency response of intra-polarization SPM filter. Dashed lines mark the $3$~dB bandwidth for the filter.}
\label{fig:blocklength}
\end{figure}

Figure~\ref{fig:blocklength} shows the magnitude Fourier transform $F_{\mathrm{x}}$ of the symbol-energy sequence $\left(E^{(c)}_{\mathrm{x}}(n)-\bar{E}^{(c)}_{\mathrm{x}}\right)$ in the $\mathrm{x}$-polarization  generated by PAS and for 1D mapping. We note that we subtract the  mean-energy  term as suggested in \eqref{e:deltadist}. Link and system parameters from  Table~\ref{tab:setup1_sim_param} are applied. For PAS, we consider CCDM blocklength $108$ and $300$, and ideal AS, where we directly draw amplitude symbols from the desired distribution. The curve for uniform signaling is also included as a reference. Overlaid to these curves is the magnitude frequency response of the intra-polarization SPM filter, with the axis on the right-hand side of the figure. The figure reveals several important observations that support previous findings considering effective SNR, e.g., \cite{fehenberger2019analysis}. First, ideal AS, which corresponds to practical AS such as CCDM with an infinite blocklength, causes more severe NLI than uniform signaling. Second, finite-length CCDM creates sequences with a high-pass characteristic, which becomes more pronounced as blocklength decreases. Hence, due to the lowpass  channel filter, nonlinear distortion is reduced, and shaped transmission can become more nonlinearity tolerant than uniform signaling. This is the reason for the nonlinear shaping gain obtained using PAS with short blocklengths. The exact cross-over point for when a gain will be observed depends on the filter bandwidth.

\subsection{Role of Multidimensional Mapping}
\label{s:filtermapping}

We can apply the same methodology as above to explain the role of  multidimensional mapping of shaped sequences (see Section~\ref{s:mapping}) for the  nonlinearity tolerance of PAS. However, before doing so, we consider a summary statistic called running digital sum (RDS), which was suggested in \cite{peng2020transmission}, to motivate the effect of multidimensional mapping on NLI. The RDS is based on the energy terms $2E_p^{(c')}+E_{p'\neq p}^{(c')}$ in approximation \eqref{eq:delta_e_simp}. The authors of \cite{peng2020transmission} only considered the sum $E_p^{(c')}+E_{p'\neq p}^{(c')}$, which is common for both polarizations $p\in\{\mathrm{x},\mathrm{y}\}$, and introduced the RDS as
\begin{equation}
\lrdst(t) = \sum_{n=0}^{t-1}\sum\limits_{c'\in{\cal C}}\sum\limits_{p'\in \cal{P}}\left(E_{p'}^{(c')}(n)-\bar{E}_{p'}^{(c')}\right)
\label{eq:RDSori}
\end{equation}
for $t\in\mathbb{N}$. 
We modify this definition by considering all energy terms in \eqref{eq:delta_e_simp}, i.e., we have the RDS for polarization $p$ as
\begin{equation}
\lrds(t) =\lrdst(t)+\sum_{n=0}^{t-1}\sum\limits_{c'\in{\cal C}}\left(E_{p}^{(c')}(n)-\bar{E}_{p}^{(c')}\right),
\label{eq:RDS}
\end{equation}
which is a more accurate predictor for nonlinearity tolerance as we will show.

The RDS can be interpreted as a lowpass filtered version of the symbol-energy sequence. 
A small absolute RDS would thus suggest a weaker nonlinear distortion.
For example, for CCDM with blocklength $\ell$ and 1D mapping (see Figure~\ref{1D}), we have  $\lrdst(t)=\lrds(t)=0$ for $t=k\cdot \ell$ and $k\in\mathbb{N}$, i.e., the RDS remains smaller for decreasing shaping blocklength. A similar argument applies to sphere-shaping based AS such as ESS, as most of the sequences are selected from energies near the surface of a sphere. We can extend the discussion to 2D mapping as done in \cite{peng2020transmission}. When CCDM-shaped amplitude sequences are mapped in two dimensions (see Figure~\ref{2D}), the sequence energy is spread over length $\ell/2$ and thus $\lrdst(t)=\lrds(t)=0$ for all $t = k \cdot \ell/2$, which predicts a further reduction in NLI. However, for 4D mapping (see Figure~\ref{4D}), the energy is spread over both polarizations, and while $\lrdst(t)=0$ for all $t = k \cdot \ell/4$, this is not the case for $\lrds(t)$. Based on the behavior of $\lrdst(t)$, it was suggested in \cite{peng2020transmission} that PAS with 4D and 2D mapping is generally superior to PAS with 2D and 1D mapping in terms of nonlinearity tolerance, respectively. While we concur with the advantage of 2D over 1D mapping, our refined RDS metric \eqref{eq:RDS} does not permit the conclusion that 4D mapping is generally advantageous.

\begin{figure}[!t]
\centering
\includegraphics[width=0.47\textwidth]{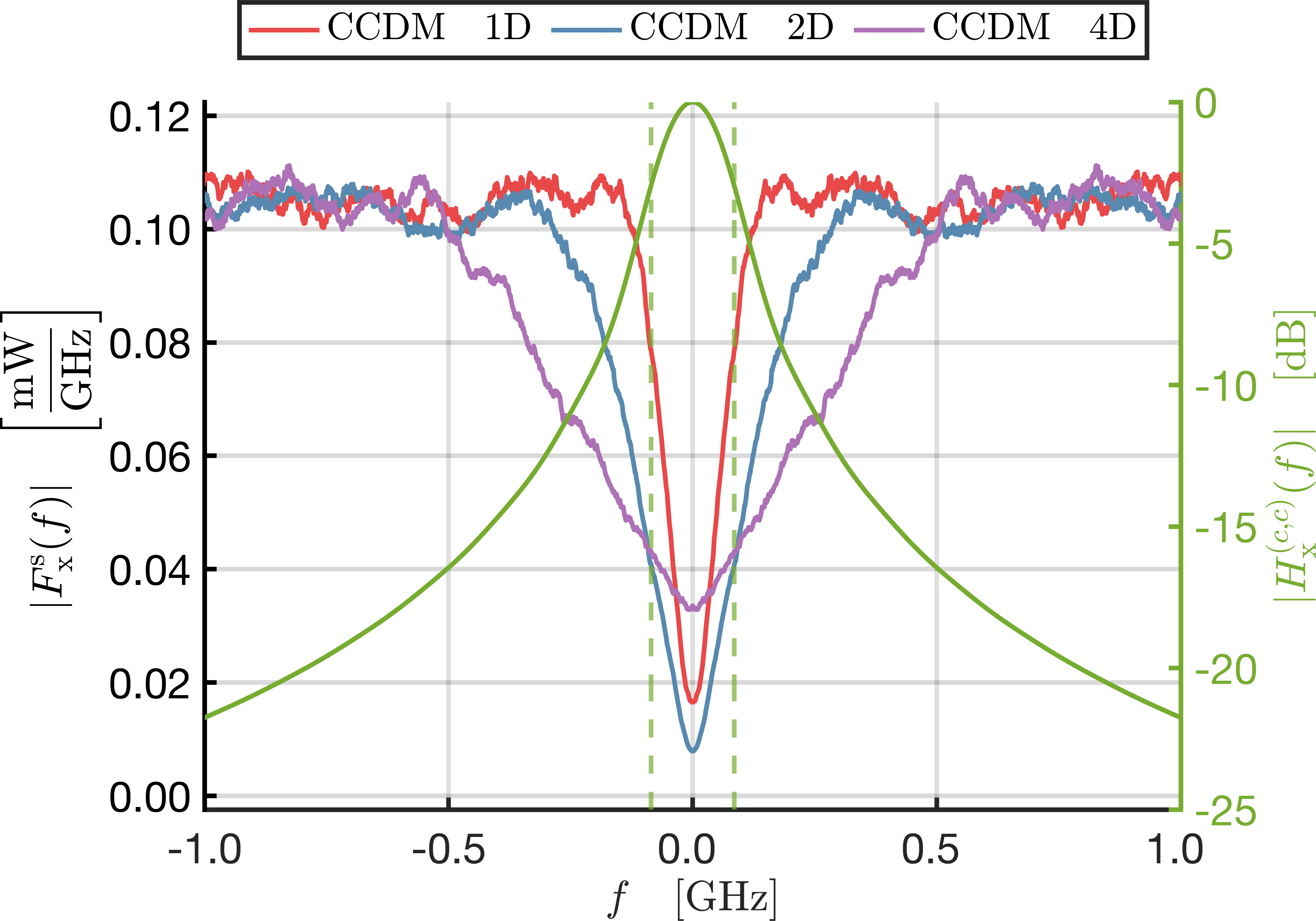}
\caption{Left: Frequency domain representation of aggregate symbol-energy sequence from \eqref{e:aggregateenergy} for ${\mathrm{x}}$-polarization generated by CCDM with blocklength $180$ and 1D, 2D, and 4D mapping. CCDM with $\ell=180$. Launch power is $-6.5$~dBm per polarization. Right: Normalized magnitude frequency response of  SPM filter $h_p^{(c,c')}$ from \eqref{eq:filter_approximation}. Dashed lines mark the $3$~dB bandwidth for the filter.}
\label{fig:mapping_dimension}
\end{figure}

We illustrate our point by plotting the absolute value of the Fourier transform $F^{\mathrm{s}}_p$ of 
\begin{equation}
\label{e:aggregateenergy}
E^{\mathrm{s}}_p(n)=2\left(E_{p}^{(c)}(n)-\bar{E}_{p}^{(c)}\right)+\left(E_{p'\neq p}^{(c)}(n)-\bar{E}_{p'\neq p}^{(c)}\right),
\end{equation}
which is the aggregate symbol-energy sequence in \eqref{eq:delta_e_simp}, for $p=\mathrm{x}$ in Figure \ref{fig:mapping_dimension}.
We consider CCDM with blocklength $180$ and 1D, 2D, and 4D mapping, and we overlay these curves with the magnitude frequency response of the inter-polarization SPM filter. Again, link and system parameters from Table~\ref{tab:setup1_sim_param} are applied. We observe that the curve for 2D mapping is always below that for 1D mapping, which means that 2D mapping always performs better than 1D mapping with respect to NLI. The comparison with 4D mapping is not that simple, as its curve intersects with those for 1D and 2D mapping. It has a shallower but wider dip around zero frequency than the 1D and 2D mapping curves. Hence, 4D mapping does not necessarily result in a reduced (or increased) NLI compared to 1D and 2D mapping, but it depends on the optical fiber link, which determines  the lowpass filter curve in Figure~\ref{fig:mapping_dimension}. In particular, we expect 4D mapping to (only) provide benefits for shorter fiber links, which produce wider lowpass filters, and long shaping blocklengths, which lead to a narrower spectral dip.
This conclusion is consistent with the results in \cite{skvortcov2021huffman} and the performance results presented in Section~\ref{results}.

\section{Metric for PAS with Sequence Selection}
\label{lsas}

In this section, we utilize the linear channel model \eqref{eq:dist} to derive a metric for PAS with sequence selection as introduced in Section~\ref{s:sequenceselection}. 

\subsection{The LSAS Metric}
An effective selection of shaped sequences suppresses the distortion variations defined in \eqref{e:deltadist}. Hence, the proposed metric is 
\begin{equation} \label{eq:dual_lsas}
\duallsas=\sum_{n=0}^{\ell/d-1} \sum_{p \in{\cal P} }\Big| s_p^{(c)}(n)\sum_{p' \in{\cal P}} \sum_{c'\in{\tilde{\cal C} \subseteq C}} \Delta D_{p,p'}^{(c,c')}(n)\Big|^2,
\end{equation} 
where, we recall, $\ell$ is the blocklength of the AS and $d$ the dimensionality of the mapping. {  $D_{p,p'}^{(c,c')}$  is obtained from the convolution of the energy sequence of each candidate with the filters $h_{p,p'}^{(c,c')}$ in \eqref{e:deltadist} and only the central $\ell/d$ convolution outputs are used in \eqref{eq:dual_lsas}. This  renders $\duallsas$ independent for successive shaping blocks.} The metric accounts for nonlinearity expressed in terms of symbol amplitudes and  filtering through the fiber channel, and we established in the previous section that the filters have lowpass characteristics. Therefore, we refer to it as the lowpass filtered symbol amplitude sequence (LSAS) metric, as already used in  \cite{taha:2022} for the special case of 1D mapping. Since LSAS measures the total NLI from a multidimensional shaped amplitude sequence with length $\ell/d$, the selection will be based on the minimum LSAS metric. In relation to the analysis in Section~\ref{spectral_dip}, we expect that LSAS-based sequence selection results in a temporal structure of symbol-energy sequences with a deeper and wider spectral dip.

\begin{figure*}[!t]
\centering
\includegraphics[width=0.8\textwidth]{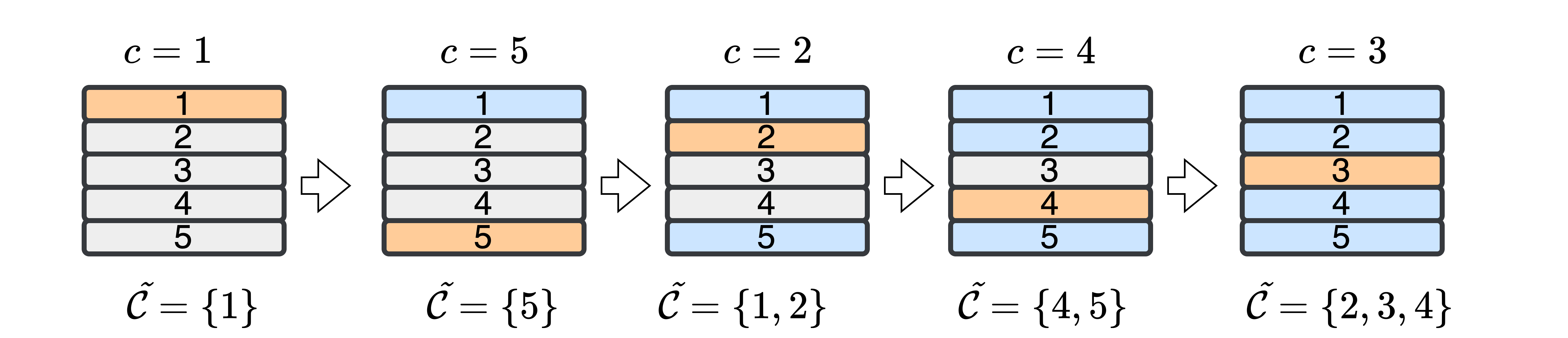}
\caption{Sequence selection for a WDM system with 5 channels ${\cal C}= \{1, 2, 3, 4, 5\}$ and  $N_{\mathrm{XPM}}=1$. $c$ and $\tilde{\cal C}$ denote the channel under selection and the subset of channels from which nonlinear effects are accounted for, respectively. The channel under selection is highlighted in orange, while the channels with selected symbols are shown in light blue.}\label{fig:xpm_greedy}
\end{figure*}

We note that we consider XPM nonlinearity from only a subset $\tilde{\cal C}$ of WDM channels in the LSAS definition \eqref{eq:dual_lsas} because the significant interference is caused by adjacent channels. 
{Ideally, a joint selection of amplitude sequences for all channels is performed. However, since then the number of candidates would grow exponentially with the number of WDM channels, we chose to perform the selection of sequences for different WDM channels sequentially, i.e., in a greedy fashion, from outer to inner channels of the system. The details of our greedy selection approach are provided in  Algorithm~\ref{alg:xpm}, for a WDM system with a set of channels $\cal C$, set of candidate sequences $\{\ve{s}^{(c,\nu)}, c \in {\cal C}, \nu\in\{1,\ldots, 2^{\frac{2|{\cal P}|}{d}v}\}\}$, and considering the XPM effects from the $N_{\mathrm{XPM}}$ closest outer channels. Fig.~\ref{fig:xpm_greedy} illustrates the execution of Algorithm~\ref{alg:xpm} for a WDM system with $|{\cal C}|=5$ channels and $N_{\mathrm{XPM}}=1$.
 }

\begin{algorithm}[t]
{ 
\scriptsize
\caption{{ Greedy LSAS sequence selection algorithm for WDM systems}} \label{alg:xpm}
\begin{algorithmic}[1]
\Require set of channels ${\cal C} =\{1,2,\dots,|{\cal C}|\}$, set of candidates $\{\ve{s}^{(c,\nu)}, c \in {\cal C}, \nu\in\{1,\ldots, 2^{\frac{2|{\cal P}|}{d}v}\}\}$, number of considered XPM channels $N_{\mathrm{XPM}}$
\Ensure selected sequences $\{\ve{\hat s}^{(c)}, c \in {\cal C}\}$

\State ${\cal T} \gets \emptyset$
\Comment{Set of channels for which selection is completed}
\For{$\alpha \in \{1,2,\dots,\frac{|{\cal C}|+1}{2}\}$}
\Comment{Loop over channels from outer to center}
\For{$\beta \in \{0,1\}$}
\Comment{For channels equidistant to the center channel}
\State $c \gets (-1)^\beta \times \alpha + \beta \times (|{\cal C}|+1)$
\Comment{Channel under selection}
\State ${\cal K} \gets \{c-N_{\mathrm{XPM}},\dots,c,\dots,c+N_{\mathrm{XPM}}\}$
\State ${\cal \tilde C} \gets ({\cal T} \cap {\cal K}) \cup \{c\}$
\State $\ve{{\hat s}}^{(c)} \gets \mathrm{candidate \ with \ minimum \ LSAS}$ {}
\State ${\cal T} \gets {\cal T} \cup \{c\}$
\EndFor
\EndFor
\end{algorithmic}}
\end{algorithm}

\subsection{Relation to the EDI Metric}
\label{subsec:edi_lsas_relation}

It is insightful to compare the LSAS metric with the EDI metric from \cite{wu2022list}. For this purpose, we show how the latter can be derived as a simplified version of the former. We specialize in the case of 1D mapping, single polarization, and single channel transmission, as EDI has been defined for this scenario. 

The EDI is computed based on the empirical average and variance of symbol-energy sequences. If we approximate the average $\bar{E}_p^{(c)}$ in \eqref{eq:expected_energy_def} with the empirical mean 
\begin{equation}
\tilde{E}_p^{(c)} = \frac{1}{\ell-w}\sum\limits_{n=1+w/2}^{\ell-w/2} |s_p^{(c)}(n)|^2, 
\end{equation}
where $w$ is the EDI window length, and introduce the average-free symbol-energy sequence 
\begin{equation}
\label{eq:delta_e_tilde}
\Delta \tilde{E}_{p,p}^{(c,c)}(n) = \sum_{m=-w/2}^{{ w/2-1}}( E_{p}^{(c)}(n+m)-\tilde{E}_p^{(c)}),
\end{equation}
then the EDI metric can be written as
\begin{equation}
\label{eq:lsas_edi}
 \ledi =\frac{1}{(\ell-w-1){w\tilde{E}_p^{(c)}}} \sum_{n=1+w/2}^{\ell-w/2} \left(\Delta \tilde{E}_{p,p}^{(c,c)}(n)\right)^2.
\end{equation}
The relation to the LSAS metric follows from noticing that the sequence $\Delta \tilde{E}_{p,p}^{(c,c)}(n)$ is an approximation of the distortion variation $\Delta D_{p,p}^{(c,c)}(n)$ from \eqref{e:deltadist}. In particular, the two quantities would be identical if $\bar{E}_p^{(c)}=\tilde{E}_p^{(c)}$ and 
\begin{equation}
\label{eq:h_edi_lsas}
h_{p,p}^{(c,c)}(n) = \begin{cases}
1,&{\text{if}}\ n\in\{-\frac{w}{2},\ldots,{\frac{w}{2}-1}\} \\ 
{0,}&{\text{otherwise}} 
\end{cases}
\end{equation}
were true.
Hence, different from the LSAS metric, the EDI in \eqref{eq:lsas_edi} does
not include the amplitude terms $|s_p^{(c)}(n)|$, uses an approximation of the lowpass filter of the channel, and it discards $w/2$ symbols at both ends of the length-$\ell$ shaped sequence to have meaningful empirical estimates. In the performance results section, we show that for these reasons, LSAS outperforms EDI for sequence selection.

\subsection{Computational Complexity Analysis}
\label{s:compcompl}
{While the main focus of our work has been on insights and improvement of NLI tolerance through shaping, we would also like to provide expressions for the associated computational cost. For this, we count the number of real-valued additions and multiplications, i.e., floating-point operations (flops), that are required for candidate generation and metric calculation.

Denoting the number of operations for sequence generation as $O_{\mathrm{gen}}$ and the number of operations for metric calculation as $O_{\mathrm{met}}$, the total number of real-valued operations for shaping with sequence selection is given by 
\begin{equation}
    \label{eq:complexity_total}
    O_{\mathrm{tot}} = \underbrace{2^{\frac{2|{\cal P}|v}{d}}}_{\mbox{\scriptsize number of} \atop {\mbox{\scriptsize candidates}}} O_{\mathrm{met}} + \underbrace{2^{v}}_{\mbox{\scriptsize number of} \atop {\mbox{\scriptsize sequences}}} O_{\mathrm{gen}}.
\end{equation} 

{The generation of each QAM sequence with length $\ell$ requires $O_{\mathrm{gen}} = 2 \times 2^{m-1}$ and $O_{\mathrm{gen}} = 4 \times 2^{m-1}$ flops per QAM symbol for ESS/K-ESS and arithmetic coding implementation of CCDM (AC-CCDM), respectively \cite{gultekin2020probabilistic}.}

The EDI metric calculation with window size $w$, candidate length $\ell$, single polarization, and 1D mapping requires 
\begin{equation}
    \label{eq:complexity_edi}
    O_{\mathrm{met}} = 6+w-\frac{w^2+2w-4}{\ell}
\end{equation}
flops per QAM symbol. We note that EDI complexity increases linearly with window size $w$ in the long blocklength regime $w \ll \ell$, where EDI is most useful.
Finally, the number of flops per QAM  symbol for evaluating the LSAS metric using filters of length $q$, mapping dimension $d$, $|{\cal P}|$ polarizations, shaping blocklength $\ell$, and $|{\cal C}|$  channels can be obtained as  
 \begin{equation}
     \label{eq:complexity_lsas}
     O_{\mathrm{met}} = 6+|{\cal P}|N_{\mathrm{avg}} - \frac{d}{|{\cal P}|\ell} + \frac{|{\cal P}| N_{\mathrm{avg}} O_{\mathrm{conv}} d}{\ell}.
 \end{equation} 
In \eqref{eq:complexity_lsas}, $O_{\mathrm{conv}}$ is the number of flops required for each convolution in \eqref{eq:dual_lsas}, and $N_{\mathrm{avg}}$ is the average number of convolutions that are evaluated for the metric calculation of different channels as 
\begin{equation}
\label{eq:avg_xpm_subchannels}
    N_{\mathrm{avg}} = \frac{2+(2N_{\mathrm{XPM}}+1)+(|{\cal C}|-3)
    (N_{\mathrm{XPM}}+1)}{|{\cal C}|
    },
\end{equation}
where $N_{\mathrm{XPM}}$ is the number of closest outer channels considered in the greedy procedure of sequence selection (see Algorithm~\ref{alg:xpm}). For simplicity, we assumed that the same filter lengths are used in all convolutions, which makes the expression in  \eqref{eq:complexity_lsas} an upper bound. We note that the LSAS complexity scales with the product $N_{\mathrm{avg}} O_{\mathrm{conv}}$, where $O_{\mathrm{conv}}$ is a monotonically increasing function of filter length $q$
when $\ell/d \ge \frac{q+1}{2}$. 

For typical settings (see Section~\ref{results}), $O_{\mathrm{gen}} \lesssim O_{\mathrm{met}}$. Since furthermore  $2^{v}\le 2^{\frac{2|{\cal P}|v}{d}}$, we will present  a quantitative comparison for $O_{\mathrm{met}}$ with EDI and LSAS in Section~\ref{subsec:results_setup1}.
} 

\section{Performance Results}
\label{results}

Sequence selection using EDI and nonlinearity tolerant amplitude shaping using K-ESS are state-of-the-art schemes for mitigating nonlinear effects in optical fiber channel. Accordingly, we consider the two simulation setups from the respective references to showcase the validity of the energy-based model and effectiveness of LSAS for sequence selection.

\begin{table}[t]
\caption{Simulation parameters for Setup~1  \cite{wu2022list}}
\label{tab:setup1_sim_param}
\centering
    \begin{tabular}{c | c}     
     \hline\hline
     Parameter & Value \\
     \hline\hline
     Modulation & 256~QAM \\
     Amplitude shaper & CCDM \\
     Shaping rate & 2.4~bits/amplitude \\
     Polarization & Single/Dual \\
     Center wavelength  & 1550~nm \\
     Symbol rate & 32~GBd \\ [1ex] 
     WDM spacing & 50~GHz \\ 
     \# WDM channels & 11 \\ 
     { Total bandwidth} & { 550~GHz} \\
     Pulse shape & Root-raised cosine\\ 
     Pulse roll-off & 0.1 \\  
     \hline
     Span length & 80~km\\
     \# Spans & 20 \\
     Fiber loss & 0.2~dB/km \\
     Dispersion parameter & 17~ps/nm/km \\
     Nonlinearity parameter ($\gamma$) & 1.37~1/W/km \\
     EDFA noise figure & 6~dB \\
     \hline
     Oversampling factor & 2 \\
     { \# QAM symbols per run} & {243,000}\\
     {\# Simulation runs} & { 5}\\
     \hline\hline 
    \end{tabular}
\end{table}

\begin{table}[t]
\caption{Simulation parameters for Setup~2 \cite{gultekin2021kurtosis}}
\label{tab:setup2_sim_param}
\centering
    \begin{tabular}{c | c}     
     \hline\hline
     Parameter & Value \\
     \hline\hline
     Modulation & 64~QAM \\
     Amplitude shaper & ESS/K-ESS \\
     Shaping rate & 1.5~bits/amplitude \\
     Polarization & Dual \\
     Center wavelength  & 1550~nm \\
     Symbol rate & 50~GBd \\ [1ex] 
     WDM spacing & 55~GHz \\ 
     \# WDM channels  & 1/11 \\
     { Total bandwidth} & { 55/605~GHz} \\
     Pulse shape & Root-raised cosine\\  
     Pulse roll-off & 0.1 \\  
     \hline
     Span length & 205~km\\
     \# Spans & 1 \\
     Fiber loss & 0.2~dB/km \\
     Dispersion parameter  & 17~ps/nm/km \\
     Nonlinearity parameter ($\gamma$) & 1.30~1/W/km \\
     EDFA noise figure & 5~dB \\
     \hline
     Oversampling factor & 2 \\
     { \# QAM symbols per run} & { 324,000}\\
     { \# Simulation runs} & { 5}\\
     \hline\hline 
    \end{tabular}
\end{table}

\subsection{Simulation Parameters}
The parameters for the two simulation setups are adopted from \cite{wu2022list} and \cite{gultekin2021kurtosis} and shown in Tables~\ref{tab:setup1_sim_param} and~\ref{tab:setup2_sim_param}, respectively.

\paragraph{Setup~1 \cite{wu2022list}}
A multi-span WDM transmission with 11 channels, 32~GBd baud rate, and 50~GHz channel spacing is simulated using the SSFM. { Two times oversampling is applied to account for fiber nonlinearity.} The modulation is 256~QAM with root raised cosine pulse shaping (roll-off 0.1). Amplitudes are generated using CCDM with a Maxwell-Boltzmann distribution to achieve an effective rate of 2.4~bits/amplitude{, i.e., the extra rate loss due to sequence selection is compensated for by increasing the AS shaping rate.}. The PAS scheme uses a  low-density parity-check (LDPC) code with rate 4/5. The overall rate is thus $5.2$~bit/QAM-symbol. The symbols are transmitted through 20 spans of a standard single-mode fiber with 80~km span length, fiber loss 0.2~dB/km, chromatic dispersion parameter 17~ps/nm/km, and nonlinearity parameter 1.37~W$^{-1}$km$^{-1}$. At the end of each span, an erbium doped amplifier (EDFA) with a 6~dB noise figure is deployed. Both single and dual polarization transmissions are considered. At the receiver, chromatic dispersion is compensated, and a mean phase rotation is applied { as a CPR, which is the same as used in \cite{wu2022list} and \cite{gultekin2021kurtosis}}. The central channel is the channel of interest for performance results. {Results are obtained as the average from  transmitting 243,000 QAM symbols in five independent simulation runs.} 

\paragraph{Setup~2 \cite{gultekin2021kurtosis}}
The main differences from Setup~1 are the adoption of 64~QAM transmission, PAS with ESS and K-ESS with a shaping rate of 1.5~bits/amplitude and a rate-5/6 FEC, i.e., the overall rate is $4$~bit/QAM-symbol, a baud rate of 50~GBd, a channel spacing of 55~GHz for the WDM case, and a single-span link of 205~km.
{ We note that there are several pairs of maximum energy and kurtosis that result in K-ESS with the same shaping rate. We used the optimal pair for blocklength $\ell=108$ as reported in \cite{gultekin2021kurtosis} for a fair comparison. As in \cite{gultekin2021kurtosis}, we will also consider 64~QAM transmission with uniform signaling and a rate 2/3-FEC as a baseline to compare various shaped transmission systems.
}

The other parameters are identical or similar to Setup~1 (see Tables~\ref{tab:setup1_sim_param} and~\ref{tab:setup2_sim_param}).

\subsection{Performance metrics}
We use three common performance metrics for assessing the nonlinearity tolerant schemes. The first is the effective SNR of symbols after linear equalization:
\begin{equation}
    \mathrm{{SNR}_{eff}} = \frac{\mathbb{E}\left[|S|^2\right]}{\mathbb{E}\left[|S-R|^2\right]},
\end{equation}
where $S$ and $R$ denote transmitted and received symbol, respectively. The second is the Q-factor, which is defined based on uncoded bit error rate (BER) as
\begin{equation}
    \mathrm{Q} = \sqrt{2} \mathrm{erfc}^{-1} (2\mathrm{BER}),
\end{equation}
where $\mathrm{erfc}$ denotes the complementary error function. The third metric is the AIR \cite{alvarado2017achievable,fehenberger2016probabilistic,fehenberger2018multiset} \begin{equation}
    \mathrm{AIR} = H(P_{\mathrm{s}}) - \sum_{i=1}^{2m} H(P_{\mathrm{b|r},i}) - 2R_{\mathrm{loss},v},
\end{equation}
where $P_{\mathrm{s}}$ and $P_{\mathrm{b|r},i}$ denote the distributions of the transmitted QAM symbol and its $i$th bit label given the received symbol, respectively. The effective SNR is useful to highlight the strength of NLI associated with different shaping schemes. Q-factor and AIR also account for the linear shaping gain. 

\subsection{Setup~1: 256~QAM Transmission}
\label{subsec:results_setup1}
We start with 256~QAM transmission in Setup~1 to compare the effectiveness of the LSAS metric for sequence selection with that of the EDI metric from \cite{wu2022list}.

\paragraph{Sequence Selection}
We first focus on sequence selection with 1D mapping. If not stated otherwise, single-polarization transmission is used. For SNR and AIR results, a launch-power optimization has been performed for each measurement point.

\begin{figure}[!t]
\centering
\includegraphics[width=0.47\textwidth]{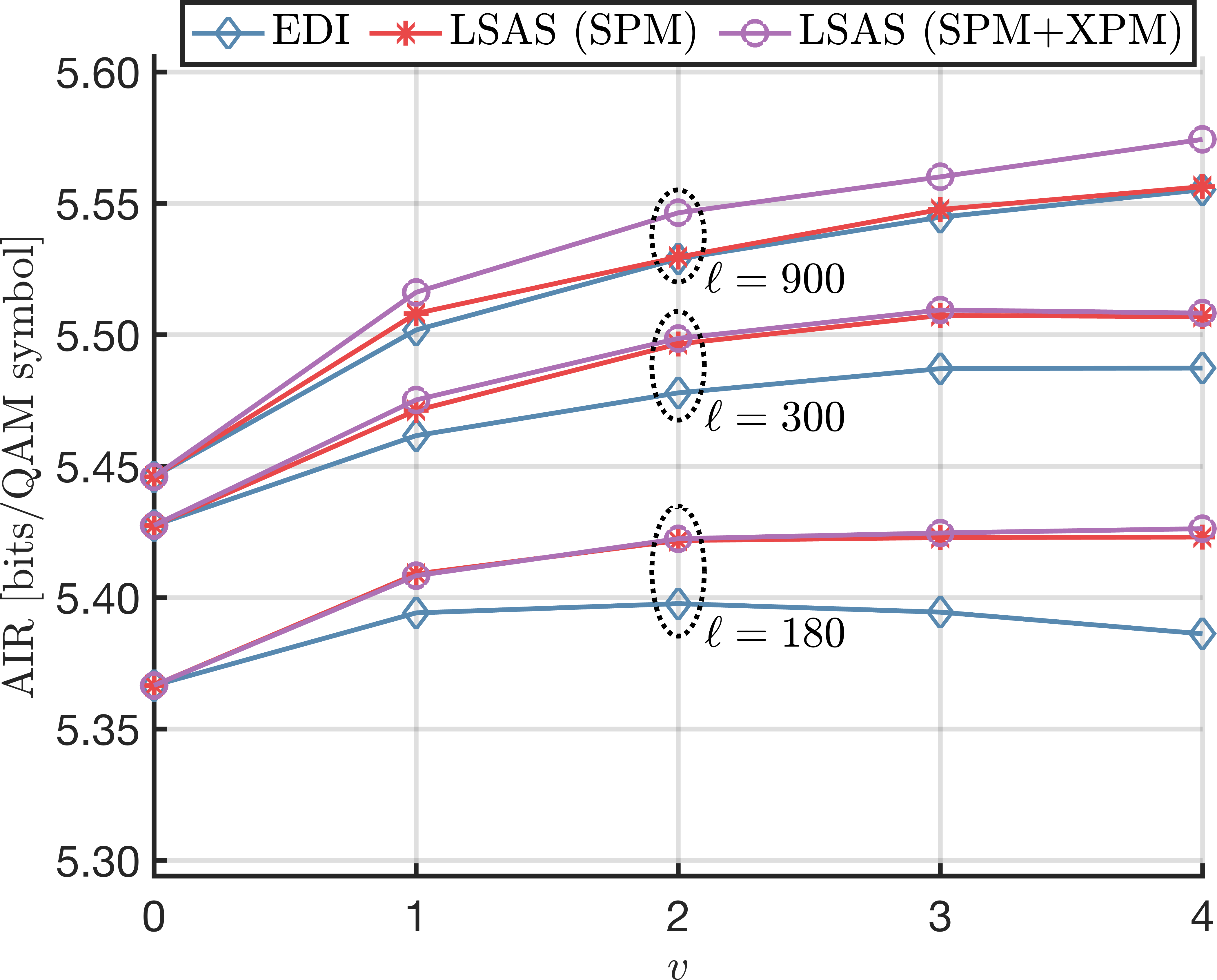}
\caption{AIR vs. $v$ for sequence selection using EDI, LSAS with only SPM components, and LSAS with both SPM and XPM components. Setup~1 with single-polarization transmission and 1D mapping.}
\label{fig:AIR_vs_v}
\end{figure}

Figure~\ref{fig:AIR_vs_v} shows the AIR as a function of the number of flipping bits $v$ for PAS with different blocklengths and EDI and LSAS metrics for sequence selection. EDI uses a window length of $w=100$, as suggested in \cite{wu2022list}. For the LSAS metric, we distinguish two cases: (i) ${\cal \tilde C} = \{ c \}$, i.e., only SPM is considered (``LSAS (SPM)"), and (ii) ${\cal \tilde C} = \{ c-1, c, c+1 \}$, i.e., XPM from two adjacent channels is considered (``LSAS (SPM+XPM)") for the center channel of interest. We observe that LSAS is superior to EDI for all blocklengths. For $\ell\in\{180,300\}$, we suspect that discarding $w$ symbols from each shaping block as well as the lowpass-filter approximation affect the EDI metric. Accounting for XPM does not make a notable difference for short blocklengths, which is due to the wide spectral dip for CCDM at $\ell\in\{180,300\}$ (see Figure~\ref{fig:blocklength}) and the narrow XPM filter (see Figure~\ref{fig:h_vs_f_b}). For $\ell=900$, the LSAS is superior to the EDI metric because of the inclusion of XPM in the selection. { Since this requires knowledge of the data in neighboring channels, it could be impractical in a conventional WDM scenario. However, it would be a legitimate option for a digital subcarrier multiplexing (DSCM) system. }
Furthermore, we note that there is an optimum value for $v$, which shifts toward larger $v$ for larger blocklengths. This is because of the trade-off between linear and nonlinear shaping gain in the sequence selection scheme. An initially higher linear shaping gain in the case of larger blocklength  provides more opportunity for increasing the nonlinear shaping gain using more redundant bits. 
However, since the computational complexity of sequence selection grows with $v$, we choose $v=2$ for the following results.

\begin{figure*}[tb]
\centering
\subfloat[]{\includegraphics[width=0.47\textwidth]{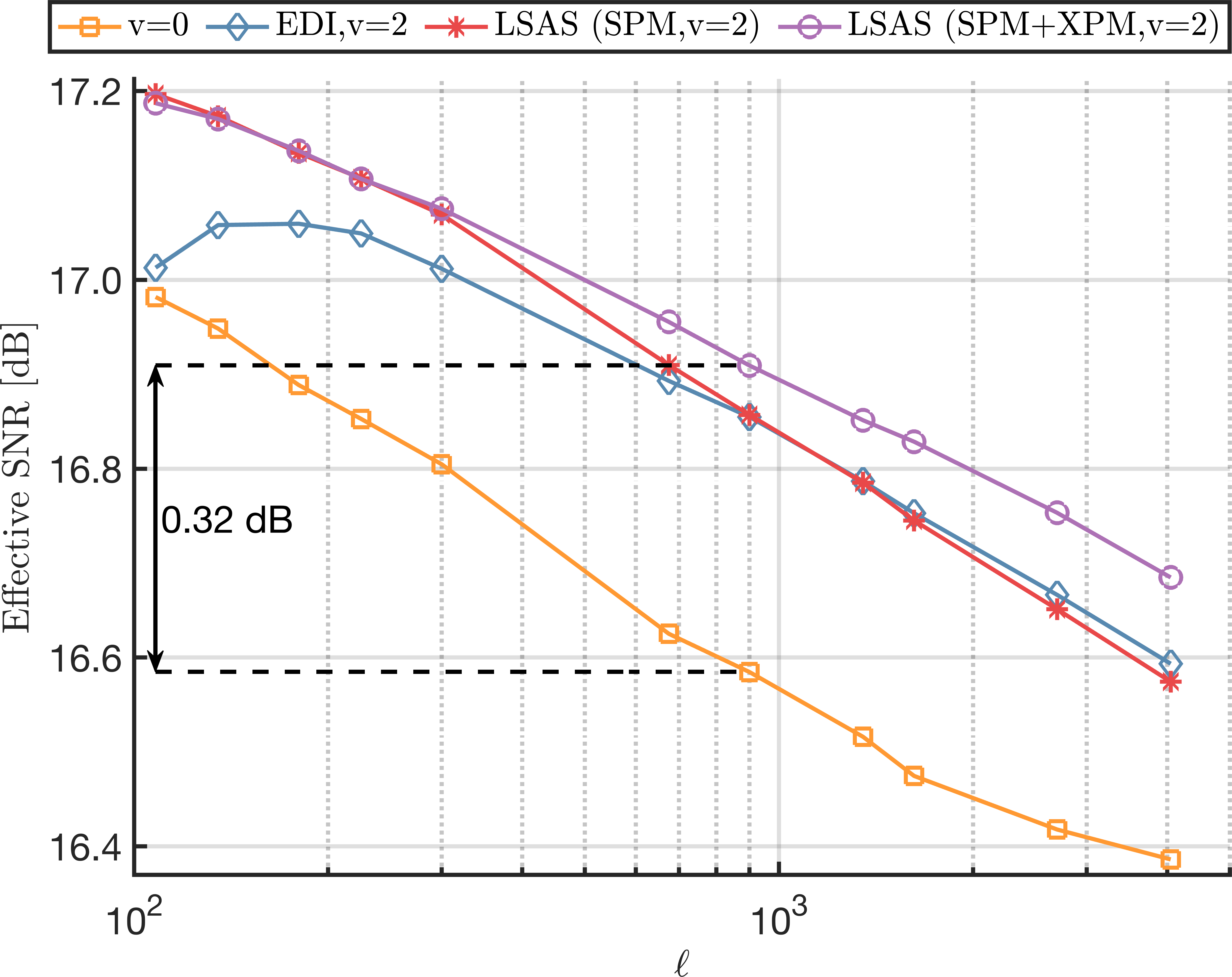}%
\label{fig:AIR_SNR_vs_L_a}}
\hfil
\subfloat[]{\includegraphics[width=0.47\textwidth]{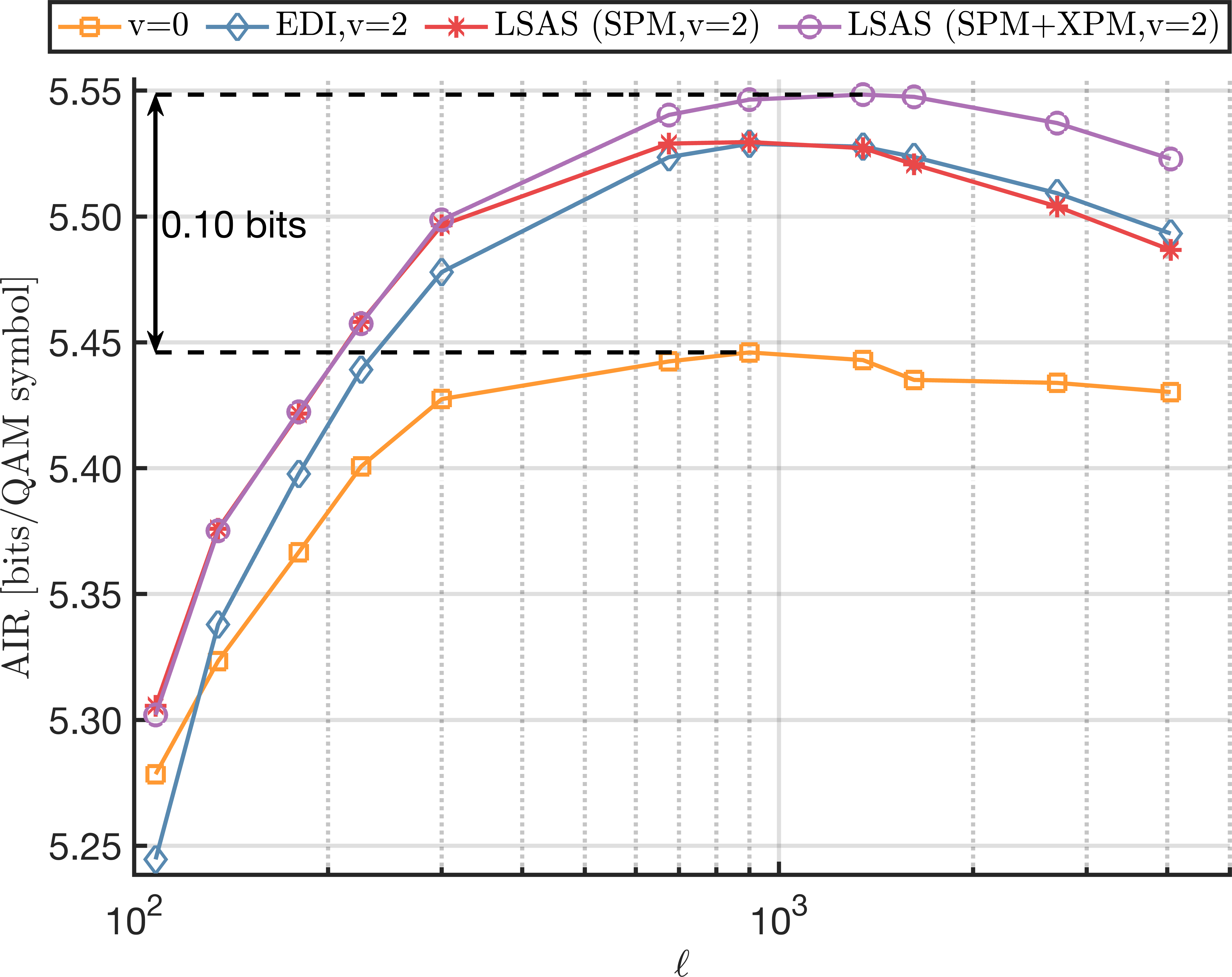}%
\label{fig:AIR_SNR_vs_L_b}}
\caption{(a) Effective SNR and (b) AIR vs.\ shaping blocklength for different selection metrics with $v=2$ redundant bits. Shaping without selection ($v=0$) shown as a reference. Setup~1 with single-polarization transmission and 1D mapping.}
\label{fig:AIR_SNR_vs_L}
\end{figure*}

Figure~\ref{fig:AIR_SNR_vs_L} compares the sequence selection metrics from Figure~\ref{fig:AIR_vs_v} for different shaping blocklengths. Shaping without sequence selection, i.e., $v=0$ is shown as the baseline curve. Selection using EDI converges to the baseline for short blocklengths, as it requires a blocklength $\ell$ notably larger than the window size $w$ to work effectively (see Section~\ref{subsec:edi_lsas_relation}). As blocklength increases, EDI and LSAS with only SPM converge to the same performance. As already noted in Figure~\ref{fig:AIR_vs_v}, LSAS with SPM and XPM components outperforms the other selection metrics for larger $\ell$, with a gain of about 0.3~dB in SNR and 0.1~bits/QAM-symbol in AIR over shaping without selection at $\ell=900$. From the AIR in Figure~\ref{fig:AIR_SNR_vs_L_b} we note that the optimal shaping blocklength  increases from $\ell=900$ in the case of no selection to $\ell=1350$ for selection using LSAS, which is due to the reduced nonlinear impairments.

{
To gauge the computational complexity associated with the two sequence selection metrics, we evaluate the expressions \eqref{eq:complexity_edi} and \eqref{eq:complexity_lsas} in Section~\ref{s:compcompl} for two shaping blocklengths $\ell=108$ and $\ell=900$ used in Figure~\ref{fig:AIR_SNR_vs_L}. The LSAS metric uses SPM and XPM filters of lengths 400 and 500, respectively, as shown in Figure~\ref{fig:h_vs_n_f}. For these values, the evaluation of EDI and LSAS (SPM) require 11 and 220 flops per QAM symbol, respectively.
 At the blocklength $\ell=900$, we have 95, 719, and 1367 flops per QAM symbol for EDI, LSAS (SPM), and LSAS (SPM+XPM), respectively, i.e., incorporating the XPM terms from the first nearest outer channels increases the computational complexity by a factor of $N_{\mathrm{avg}} \approx 2$. We thus observe a significantly increased computational complexity for LSAS compared to EDI, which is due to the scaling of complexity with filter length and $N_{\mathrm{avg}}$. 
However, the LSAS filter lengths can be shortened with little effect on the performance results. 
For example, we can simply truncate the LSAS filters using  a threshold compared to the maximum coefficient. Table~\ref{tab:complexity} shows the number of flops and the corresponding AIR for different truncation thresholds.
We observe that the computational complexity of LSAS can be reduced to values almost identical to those for EDI, i.e., 16 vs.\ 11 flops at a threshold of $-8$~dB for $\ell=108$ and 87 vs.\ 95 flops at a threshold of $-15$~dB for $\ell=900$, with a negligible loss in AIR. We speculate that this can further be improved with more sophisticated filter approximation approaches.}

In the remainder of this section, results for LSAS refer to LSAS with only SPM components, which is suitable for blocklengths $\ell\le 300$.

\begin{table}[t]
{
\caption{{ Number of real-valued operations per QAM symbol and AIR  (in [bits/QAM symbol]) for LSAS (SPM) at $\ell=108$ and $\ell=900$ for different truncation thresholds (in dB).}}\vspace*{-3mm}
\label{tab:complexity}
\begin{center}
    \begin{tabular}{|c|| c | c | c | c | c|}     
     \hline
     $\ell$ & Threshold & $-\infty$ & $-20$ & $-15$  & $-8$ \\
     \hline\hline 
     \multirow{2}{*}{{$108$}}& $O_{\mathrm{LSAS}}$ & 220 & 145 & 80 & 16 \\
     & AIR  & 5.30 & 5.30 & 5.30 & 5.29 \\
     \hline
     \multirow{2}{*}{{$900$}}& $O_{\mathrm{LSAS}}$ & 719  & 176 & 87 & \\
     & AIR  & 5.53 & 5.53 & 5.52 & \\
     \hline
    \end{tabular}
    \end{center}
    }
\end{table}

\begin{figure}[!t]
\centering
\includegraphics[width=0.47\textwidth]{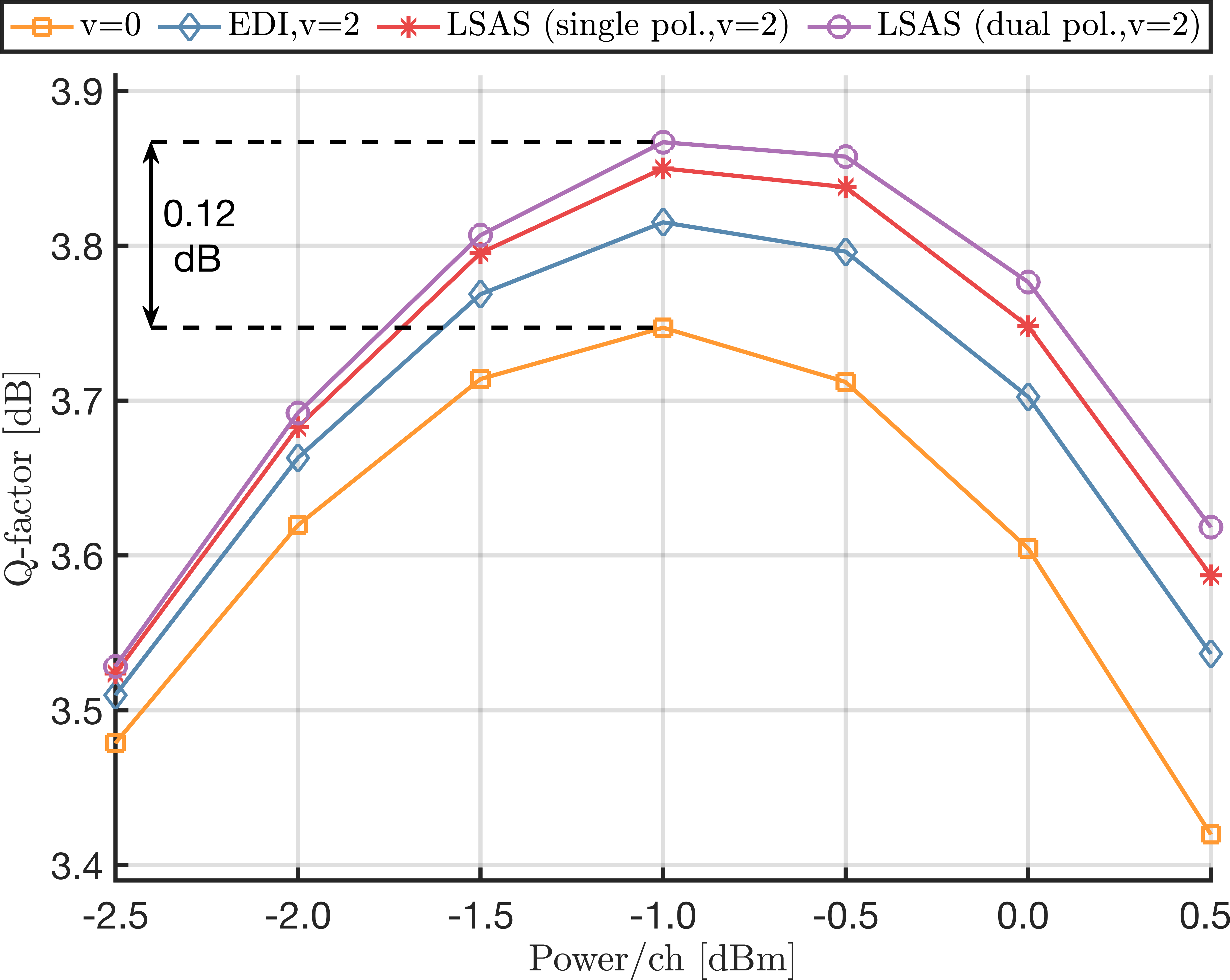}
\caption{Q-factor vs.\ launch power for different selection metrics with $v=2$ redundant bits. Shaping without selection ($v=0$) shown as a reference. Setup~1 with dual-polarization transmission, 1D mapping, and shaping blocklength $\ell=180$.}
\label{fig:Q_factor_vs_p}
\end{figure}

\begin{figure}[!t]
\centering
\includegraphics[width=0.47\textwidth]{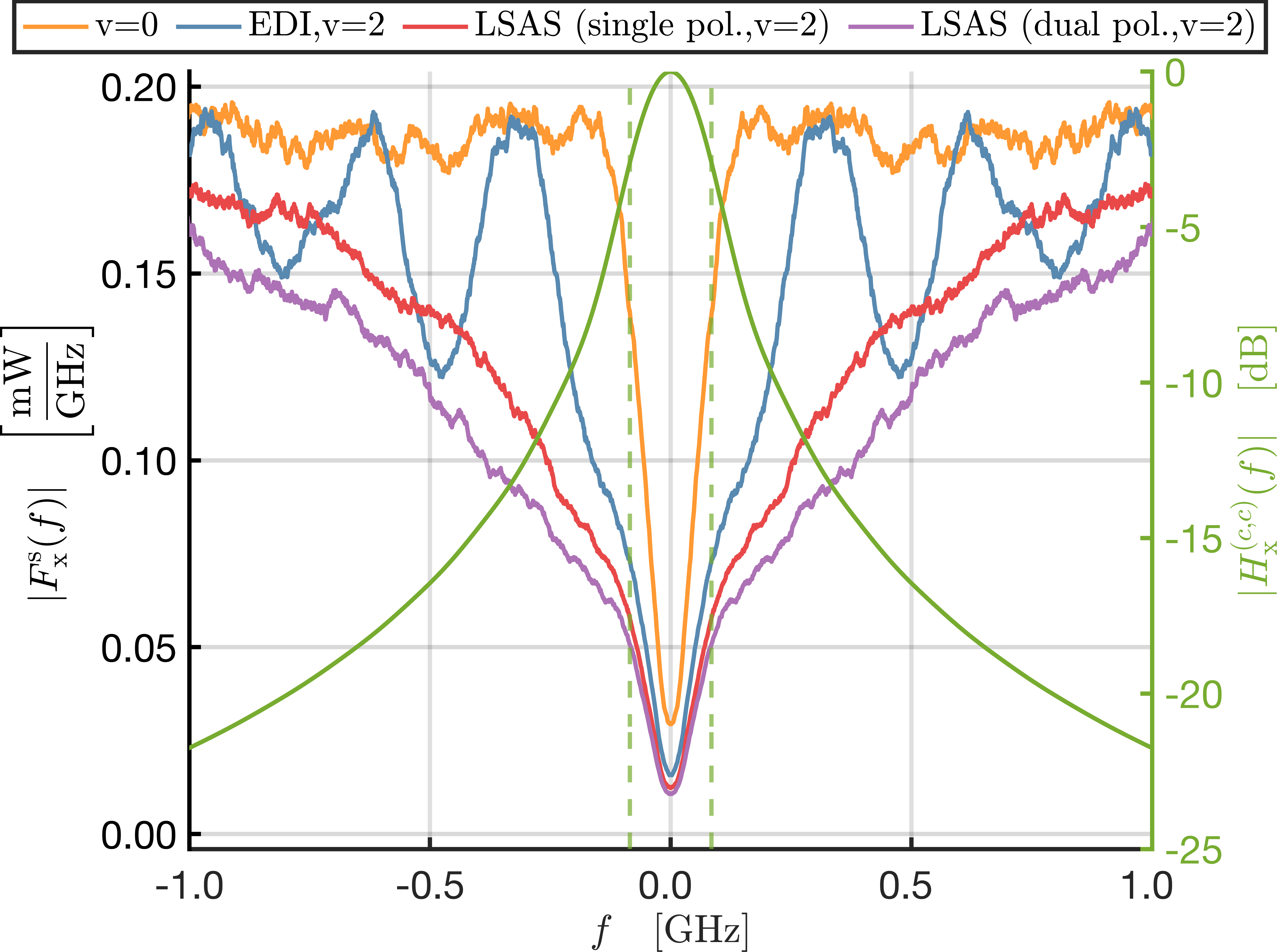}
\caption{Left: Frequency domain representation of aggregate symbol-energy sequence from \eqref{e:aggregateenergy} for $\mathrm{x}$-polarization for shaping methods and system setup from Figure~\ref{fig:Q_factor_vs_p}.
Launch power is $-1$~dBm per channel. Right: Normalized magnitude frequency response of SPM filter $h_p^{(c,c')}$ from \eqref{eq:filter_approximation}. Dashed lines mark the $3$~dB bandwidth for the filter.}
\label{fig:selection_frequency_analysis}
\end{figure}

Next, we present results for dual-polarization transmission.  Figure~\ref{fig:Q_factor_vs_p} shows the Q-factor as a function of  launch power for dual-polarization transmission and shaping with different selection metrics. In the case of the LSAS metric, we distinguish between $({\cal P}=\{{\mathrm{x}}\}, {\cal P}=\{{\mathrm{y}}\})$ and ${\cal P}=\{{\mathrm{x},{\mathrm{y}}}\}$ in \eqref{eq:dual_lsas}, i.e., a single-polarization and a dual-polarization version. We observe that the advantage of the LSAS metric also manifests in the dual-polarization case, with some slight additional gain due to performing joint selection over the two polarization dimensions. We highlight the effect of shaping sequence selection and the different selection metrics in Figure~\ref{fig:selection_frequency_analysis}, which shows the Fourier transforms $F_p^{\mathrm{s}}$ of the aggregate symbol-energy sequences defined in \eqref{e:aggregateenergy} corresponding to the different shaping methods. We overlay the curves with the frequency response for the SPM filter $h_p^{(c,c')}$ defined in \eqref{eq:filter_approximation}, which provides the approximation for the distortion in \eqref{eq:delta_e_simp}.
It can be seen that sequence selection makes the spectral dip wider and deeper, which translates to a decrease in NLI considering the lowpass channel filter. Moreover, the order of the curves in Figure~\ref{fig:selection_frequency_analysis} matches the order of the corresponding curves in Figure~\ref{fig:Q_factor_vs_p}. This again demonstrates the usefulness of the linear channel model for nonlinearity analysis. As a final detailed observation, we point out the oscillation of the Fourier transform signal for the EDI metric in Figure~\ref{fig:selection_frequency_analysis}. This results from rectangular-filter approximation in  \eqref{eq:h_edi_lsas}, which corresponds to a sinc-function in the frequency domain, and thus frequency components at its nulls do not contribute to the metric.

\begin{figure}[!t]
\centering
\includegraphics[width=0.47\textwidth]{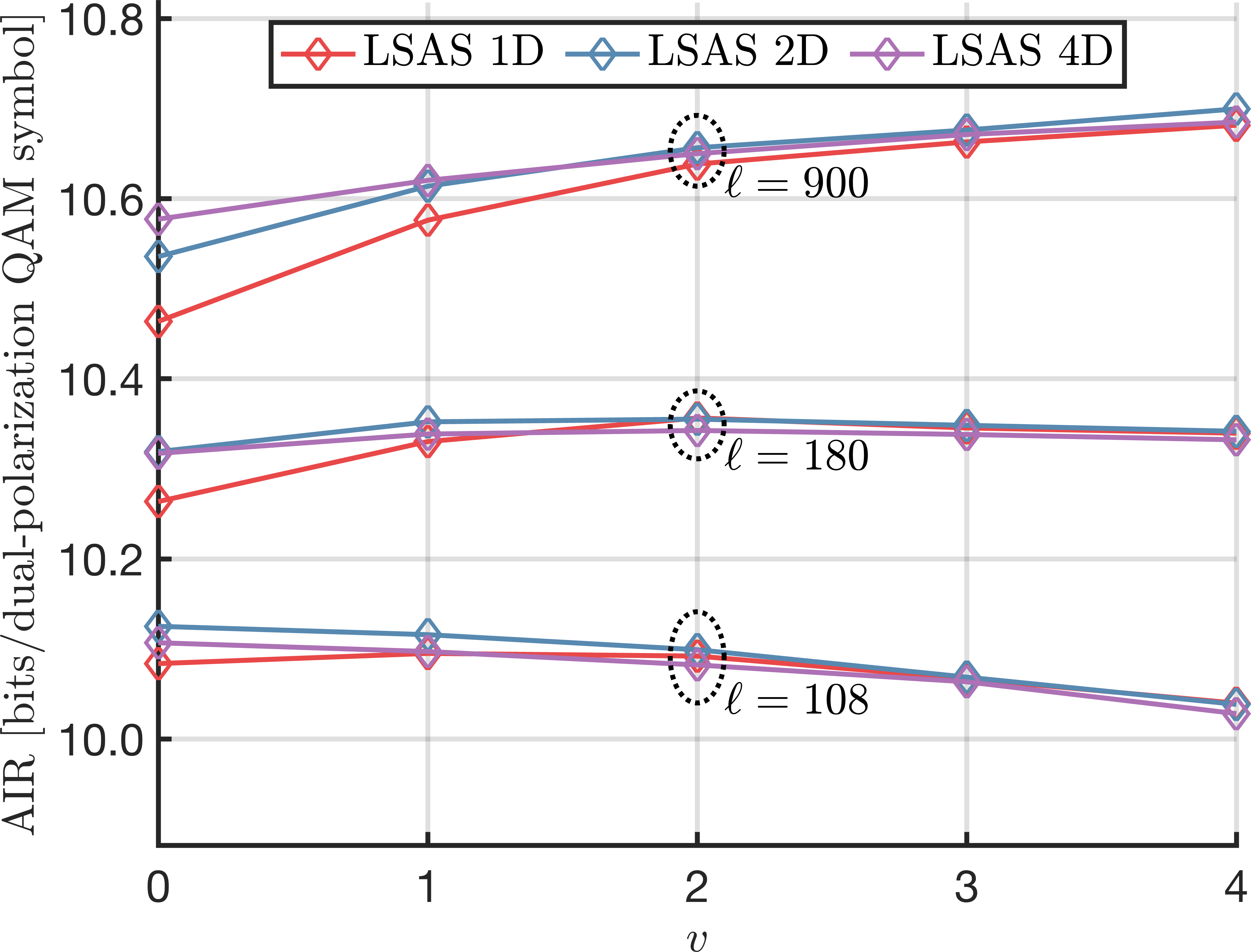}%
\caption{AIR vs.\ $v$ for sequence selection with LSAS and different mapping strategies. Setup~1 with dual-polarization transmission.
}
\label{fig:AIR_SNR_MD_mapping}
\end{figure}

\paragraph{Multidimensional mapping}
As the last set of results for Setup~1, we consider sequence selection with different amplitude mapping strategies. For this, we need to adopt the dual-polarization transmission scenario. Furthermore, since the  components of the 4D signal need to be chosen jointly, only the dual-polarization version of the LSAS metric but not the EDI can be applied. Figure~\ref{fig:AIR_SNR_MD_mapping} shows the AIR as a function of $v$ for the three mapping strategies from Section~\ref{s:mapping}. First, we observe that for shaping  without selection ($v=0$), the dimensionality of mapping has a notable effect on the system performance. 4D mapping is superior to 1D and 2D mapping strategies only for long blocklengths. This result has been predicted in Section~\ref{s:filtermapping} from the channel model for distortion and the frequency-domain visualization in Figures~\ref{fig:blocklength} and \ref{fig:mapping_dimension}. 
{ Second, as the number of selection bits increases, the performances for all three mapping  strategies become identical even though they select among different number of candidates.} Therefore, it is preferable to use 4D mapping in conjunction with sequence selection for its lower complexity.
 { Regarding the latter, one could also compare the different mapping strategies for the same number of candidates instead of fixed shaping blocklength $\ell$. 
 For example, denoting the number of redundant bits for mapping dimension $d$ as $v_d$, then we can compare  $v_1=1$ from the 1D mapping curve with $v_2=2$ and $v_4=4$ from the 2D and 4D mapping curves in  Figure~\ref{fig:AIR_SNR_MD_mapping}, respectively. From this comparison  we observe that multidimensional mapping is superior to 1D mapping for relatively longer blocklengths, i.e., $\ell=900$ in Figure~\ref{fig:AIR_SNR_MD_mapping}. For shorter blocklengths, it is preferable to use a smaller number of candidates to avoid the larger rate loss with increasing $v$.}

\subsection{Setup~2: 64~QAM transmission}

\begin{figure}[!t]
\centering
\includegraphics[width=0.47\textwidth]{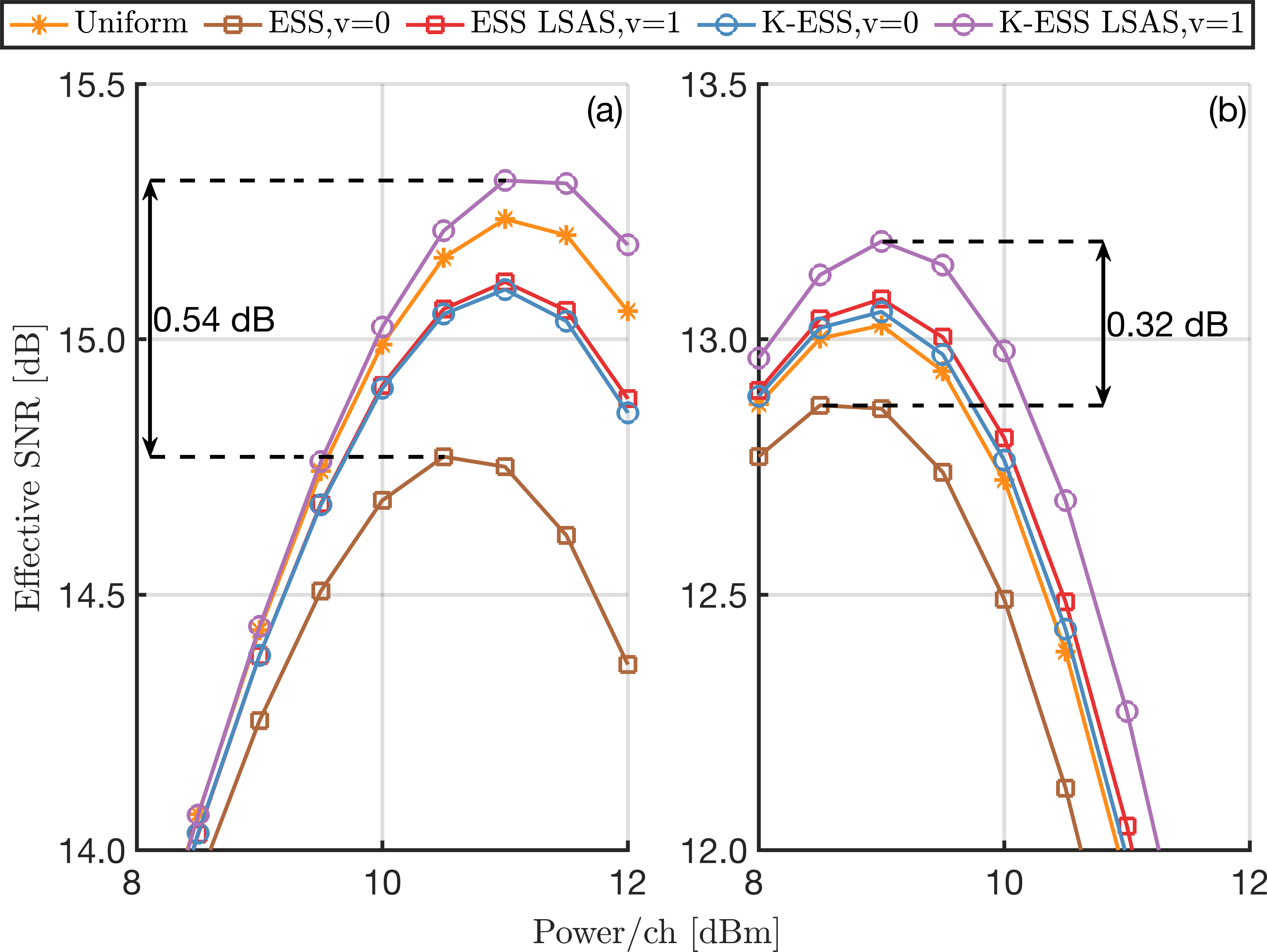}
\caption{Effective SNR vs.\ launch power for Setup~2. (a) Single channel  transmission and (b) WDM transmission with 11 channels. Shaping blocklength $\ell=108$ and 4D mapping.} 
\label{fig:SNR_KESS_single_span}
\end{figure}
For Setup~2 from \cite{gultekin2021kurtosis}, we focus on the comparison of sequence selection using LSAS with K-ESS, which was proposed for improved nonlinearity tolerance in \cite{gultekin2021kurtosis}, as well as for the first time, the combination of the two methods. 

\begin{figure}[!t]
\centering
\includegraphics[width=0.47\textwidth]{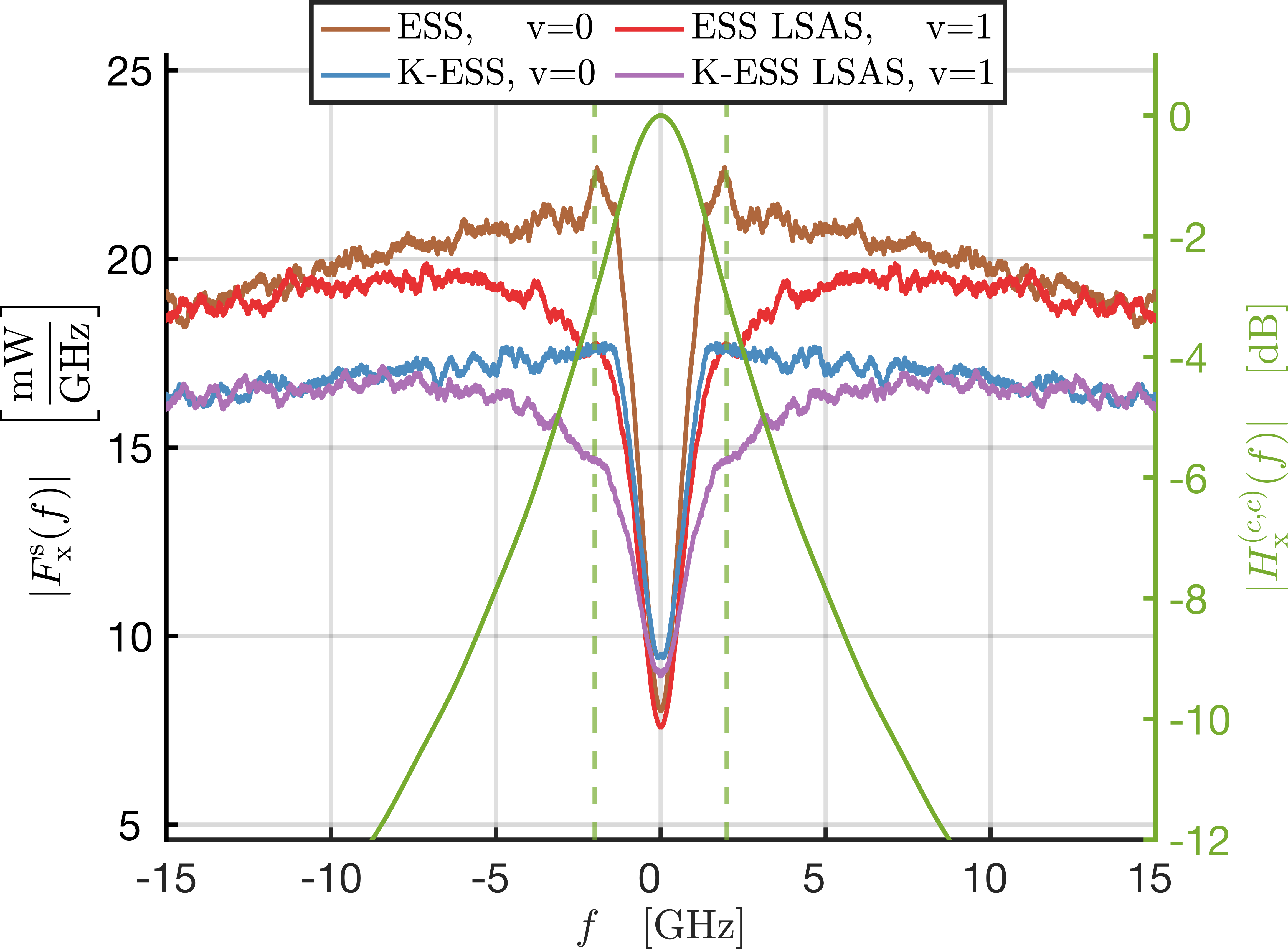}
\caption{Left: Frequency domain representation of aggregate symbol-energy sequence from \eqref{e:aggregateenergy} for $\mathrm{x}$-polarization for shaping methods from Figure~\ref{fig:SNR_KESS_single_span}(b).
Launch power is $9$~dBm per channel. Right: Normalized magnitude frequency response of  SPM filter $h_p^{(c,c')}$ from \eqref{eq:filter_approximation}. Dashed lines mark the $3$~dB bandwidth for the filter. 
} 
\label{fig:kess_freq}
\end{figure}

Figure~\ref{fig:SNR_KESS_single_span} shows the effective SNR for the various shaping schemes for (a) single channel  and (b) WDM transmission. PAS with $\ell=108$ and 4D mapping are applied, where the latter is beneficial because of the short link length. The effective SNR for uniform signaling is included as a reference. We observe that PAS with ESS suffers an SNR loss compared to uniform signaling, which is due to an increased NLI. A good fraction of this loss (single-channel case) or even the entire loss (WDM case) is recovered using K-ESS or adding sequence selection with LSAS. The combination of K-ESS with LSAS-based selection outperforms all methods with a gain of more than 0.5~dB and 0.3~dB over ESS-based shaping in the single-channel and WDM cases, respectively.   
The SNR-performance results can be explained by considering the frequency-domain representation of the corresponding symbol-energy signals and  the linear channel filter in Figure~\ref{fig:kess_freq}. It can be seen that K-ESS is effective in lowering the higher-frequency components, which is a result of generating a distribution with lower kurtosis. The selection-based schemes optimize the temporal structure of symbols to minimize the NLI with the same distribution, which translates to a wider and deeper spectral dip at frequency zero. As a result, we expect that the selection-based schemes perform well for a wide variety of transmission setups, while K-ESS is only beneficial for short fiber links corresponding to a wide lowpass filter. This conclusion is consistent with the observations in  \cite{gultekin2021kurtosis}.

\begin{figure}[!t]
\centering
\includegraphics[width=0.47\textwidth]{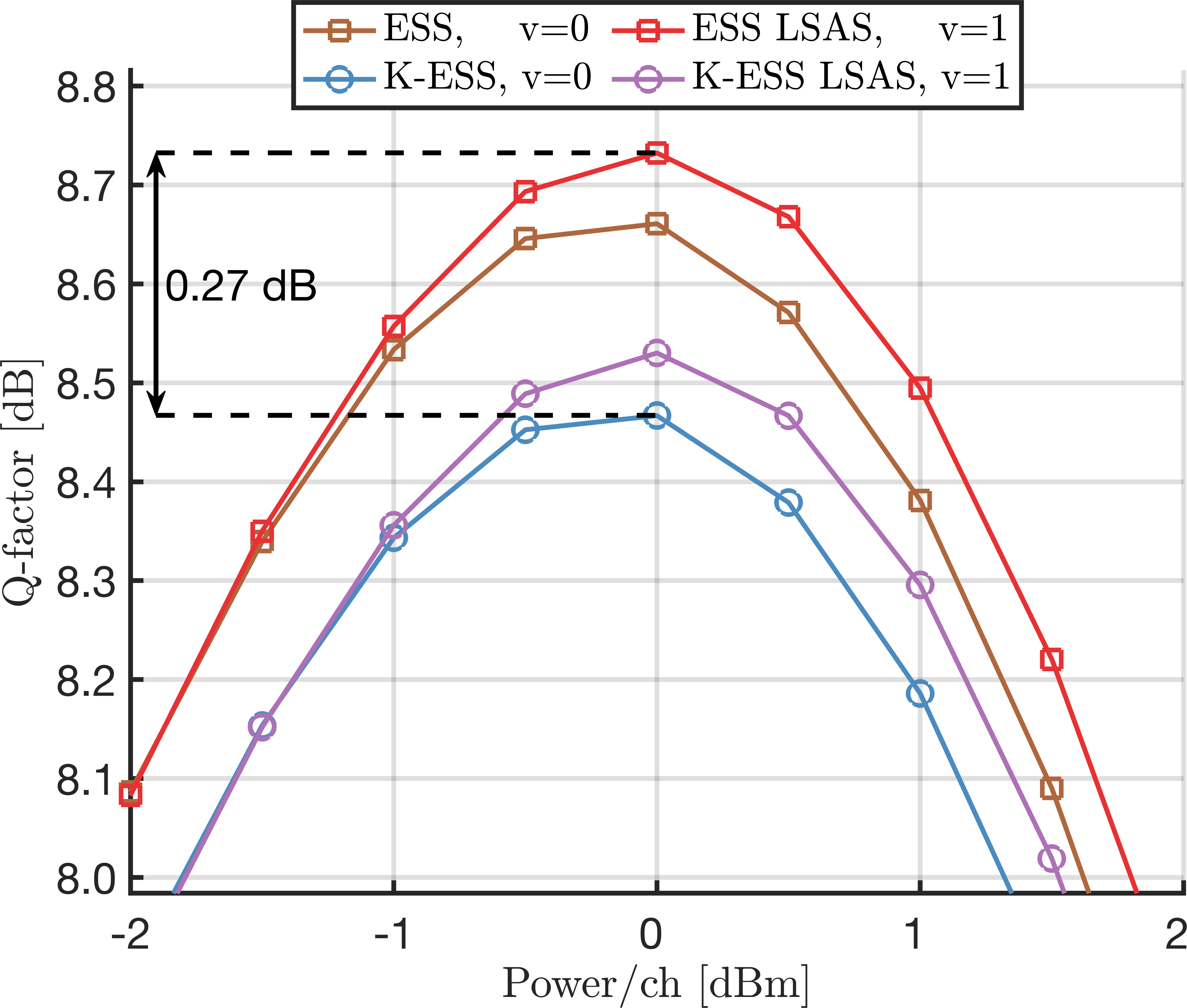}
\caption{Q-factor vs.\ launch power for Setup~2 but with a single channel transmission through $20$ spans of $80$~km. Shaping blocklength $\ell=108$ and 2D mapping strategy.} 
\label{fig:Q_factor_KESS_multi_span}
\end{figure}

Finally, to examine our hypothesis about K-ESS, we consider Setup~2 but with a single channel transmission through a long-haul fiber link with $20$ spans of length $80$~km. Figure~\ref{fig:Q_factor_KESS_multi_span} shows the Q-factor curves obtained with ESS and K-ESS with $\ell=108$ and with and without LSAS-based sequence selection. 2D mapping is chosen as it is the best mapping strategy for this longer link. As suggested by our linear channel model analysis, ESS outperforms K-ESS in this scenario. Furthermore, sequence selection provides a gain when applied to either baseline AS.

\section{Further Discussion}
\label{sec:discussion}
{
In the derivation of the LSAS metric we applied several simplifications that permitted us to apply sequence selection for each shaping block independently. First, 
we ignored the terms in the perturbative model  \eqref{signal-signal pb} that also depend on signal phases. This is because the latter are determined by the sign bits that are generated by the FEC encoder (see Section~\ref{sec:pas}). As one FEC block typically consists of several shaping blocks, sign bits and thus signal phases are dependent on multiple shaping blocks. Second, the channel memory, i.e., the convolution of energy sequences with the filters $h_{p,p'}^{(c,c')}$ in \eqref{e:deltadist}, creates dependencies between NLI of adjacent shaping blocks. In the LSAS metric \eqref{eq:dual_lsas} we ignored these dependencies by applying energy sequences for each shaping block separately to the convolution operation. Despite these simplifications, our results in Section~\ref{results} showed that the proposed linear channel model and selection metric are sensitive enough to guide the interaction between shaping and fiber nonlinearity. A natural extension of this work is a selection mechanism that accounts for sign bits and fully includes the channel memory. In this case, a joint selection over multiple shaping blocks is in order. However, a naive approach would substantially increase the computational complexity, as the number of possible sequences grows exponentially with
the number of shaping blocks. Therefore, it is interesting to explore possible alternative methods that do not suffer from the complexity growth. 

Besides the sign bits, we note that the implementation of the CPR also affects the nonlinear gain of PAS \cite{civelli2020interplay}. The proposed LSAS metric accounts for a mean phase rotation at the receiver. Similarly, the simulation setup used  in Section~\ref{results} applies such a CPR, which has been consistent with the setups in related previous works  \cite{wu2022list, gultekin2021kurtosis}. This choice is motivated as it enables us to better delineate the effects of nonlinearity tolerant shaping and different selection metrics, which is the main focus of this paper. {To further demonstrate the performance with sequence selection entirely independent of a CPR implementation, we consider the nonlinear phase noise introduced by the fiber channel based on the additive-multiplicative distortion model obtained from first-order perturbation theory \cite{liang2014multi}. We apply Setup~1 with single polarization transmission and evaluate the phase noise terms in the distortion model to account for SPM and XPM 
from the two adjacent channels on either side of the channel of interest. In Figure~\ref{fig:phase_noise}, we present the estimated standard deviation of nonlinear phase noise plotted versus the shaping blocklength for different selection metrics. The figure compares PAS without sequence selection ($v=0$), selection using EDI, and selection using LSAS. The results show that sequence selection successfully suppresses nonlinear phase noise. Similar to the effective-SNR performance metric in Figure~\ref{fig:AIR_SNR_vs_L_a}, selection using the EDI metric does not produce any nonlinear gain in the short blocklength regime. However, for longer blocklengths, all selection metrics decrease the standard deviation of nonlinear phase noise. Furthermore, the order of curves in Figure~\ref{fig:phase_noise} matches the effective-SNR result in Figure~\ref{fig:AIR_SNR_vs_L_a}. Considering the interplay of PAS and CPR for suppressing nonlinear phase noise,  a desirable extension of this work is the formulation of a sequence-selection metric that accounts for receiver processing with a more potent CPR method than a mean phase rotation.}

\begin{figure}[!t]
\centering
\includegraphics[width=0.47\textwidth]{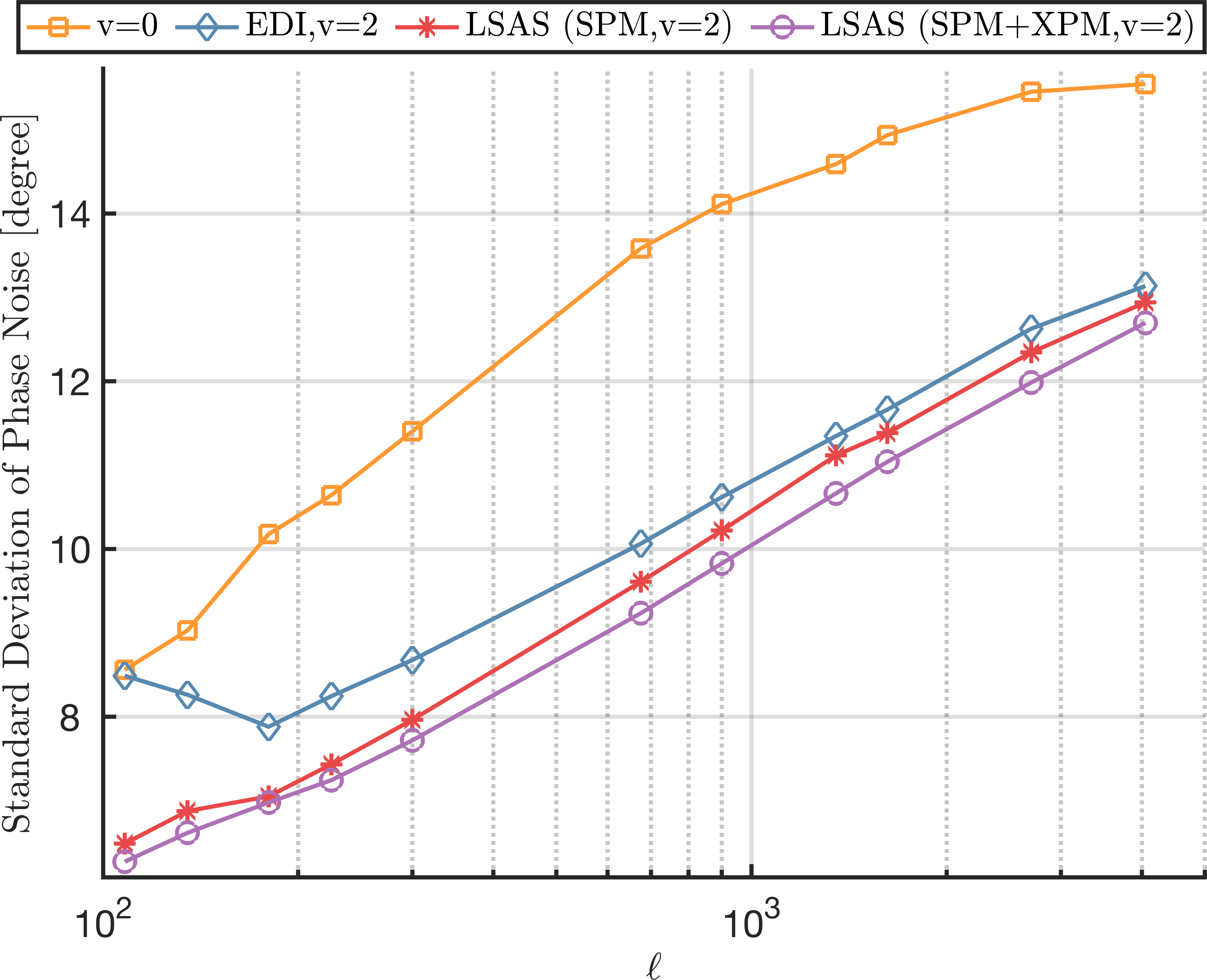}
\caption{Estimated standard deviation of nonlinear phase noise vs.\ shaping blocklength for PAS without ($v=0$) and with sequence selection ($v=2$). Sequence selection using EDI and LSAS metrics. Setup~1 with single-polarization transmission and 1D mapping.
} 
\label{fig:phase_noise}
\end{figure}

\section{Conclusion}
\label{conclusion}
In this paper, we revisited the signal-dependent generation of NLI in optical channels and presented a linear lowpass channel model to approximately describe this process. The model considers symbol-energy sequences as input and therefore permits the analysis of the interplay between amplitude shaping and nonlinear distortion. We showed the evolution of filter coefficients for different fiber setups and introduced a frequency domain visualization as a useful tool for the analysis and design of PAS schemes for nonlinearity mitigation. We applied the model to develop the LSAS metric for PAS with sequence selection. LSAS naturally includes inter-polarization and inter-channel nonlinear effects and extends to multidimensional mapping. The latter permits the combination of sequence selection with multidimensional mapping for lower complexity. Through extensive simulations, we showed the performance benefits resulting from LSAS, and we confirmed the accuracy of the predictions based on the linear model.

\setcounter{secnumdepth}{-1}
\section{Appendix}
\label{appendix_a}

In this appendix, we derive the approximation \eqref{eq:energy_model_a} for the linear channel model between symbol-energy sequences and nonlinear distortions. 

We start with the expression \cite[Eq.~(3)]{dar2014inter} for the nonlinear distortion term from signal-signal interactions based on the first-order perturbation analysis of a multi-span dual-polarization coherent WDM system. Accordingly, the input-output relationship for the $\mathrm{x}$-polarization signal in WDM channel $c \in {\cal C}$ after linear equalization can be written as 
\begin{equation} \label{signal-signal pb}
\begin{split}
 &r_\x^{(c)}(n) = s_\x^{(c)}(n) \\ 
 &+ {\j}\gamma \sum_{m \in \mathbb{Z}} \sum_{k \in \mathbb{Z}} \bigg( s_\x^{(c)}(m+n)[{s_\x^{(c)}(k+m+n)}]^* s_\x^{(c)}(k+n) \\
 &+ s_\y^{(c)}(m+n)[s_\y^{(c)}(k+m+n)]^*s_\x^{(c)}(k+n) \bigg)h_{\spm}^{(c)}(m,k) \\
 &+ {\j} \gamma \!\!\!\sum_{c'\in \atop{\cal C}\backslash\{c\}} \!\!\sum_{m \in \mathbb{Z}} \sum_{k \in \mathbb{Z}} \bigg(\!2s_\x^{(c')}(m\!+\!n)[s_\x^{(c')}(k\!+\!m\!+\!n)]^* s_\x^{(c)}(k\!+\!n) \\
 &+ s_\y^{(c')}(m+n)[s_\y^{(c')}(k+m+n)]^* s_\x^{(c)}(k+n) \\ &+ s_\x^{(c')}(m+n)[s_\y^{(c')}(k+m+n)]^* s_\y^{(c)}(k+n)\bigg)h_{\xpm}^{(c,c')}(m,k),
\end{split}
\end{equation}
where $s_{p}^{(c)}(n)$ and $r_{p}^{(c)}(n)$ are the $n$th transmitted symbol and received sample in polarization $p\in\{\mathrm{x},\mathrm{y}\}$ and channel $c \in {\cal C}$, respectively, and $\gamma$ is the fiber nonlinearity parameter. $h_{\spm}^{(c)}$ and $h_{\xpm}^{(c,c')}$ contain the perturbation coefficients for intra-channel and inter-channel NLI, respectively.
An analogous expression to \eqref{signal-signal pb} can be written for the $\mathrm{y}$-polarization signal. 

To arrive at the representation used for the distortion model based on symbol-energy sequences, we approximate  \eqref{signal-signal pb} 
by retaining only the symbol-energy dependent terms in the summations:
\begin{equation} \label{signal-signal pb-approx}
\begin{split}
 &r_\x^{(c)}(n) \approx s_\x^{(c)}(n) \Big[1+\\ 
 &\j\gamma \sum_{m \in \mathbb{Z}}  \bigg( |s_\x^{(c)}(m+n)|^2 +|s_\y^{(c)}(m+n)|^2 \bigg)h_{\spm}^{(c)}(m,0) \\
 &+\j\gamma  \sum_{k \in \mathbb{Z}\backslash\{0\}}  \bigg( |s_\x^{(c)}(k+n)|^2 \bigg)h_{\spm}^{(c)}(0,k) \\
 &+\j\gamma   \!\!\!\sum_{c'\in \atop{\cal C}\backslash\{c\}} \!\!\sum_{m \in \mathbb{Z}} \bigg(\!2|s_\x^{(c')}(m\!+\!n)|^2\!+\!|s_\y^{(c')}(m\!+\!n)|^2 \bigg)h_{\xpm}^{(c,c')}(m,0)\Big].
\end{split}
\end{equation}
Then, noting that $h_{\spm}^{(c)}(m,0)=h_{\spm}^{(c)}(0,m)$ \cite{yan2011low}, and defining filters $h^{(c,c')}_{p,p'}$ as 
\begin{equation} 
\label{eq:filter_coeff}
\begin{split}
&c'=c,p'=p: h^{(c,c')}_{p,p'}(n) =
\begin{cases}
2h_{\spm}^{(c)}(n,0), &\quad n\neq 0 \\
h_{\spm}^{(c)}(n,0), &\quad n = 0
\end{cases}\\
&c' = c,p' \neq p: h^{(c,c')}_{p,p'}(n)=h_{\spm}^{(c)}(n,0)\\
&c' \neq c,p' = p: h^{(c,c')}_{p,p'}(n)=2h_{\xpm}^{(c)}(n,0)\\
&c' \neq c,p' \neq p: h^{(c,c')}_{p,p'}(n)=h_{\xpm}^{(c)}(n,0)
\end{split}
\end{equation}
to describe both intra- and inter-channel NLI effects, we obtain \eqref{eq:energy_model_a} in Section~\ref{energy_model}.

Furthermore, as it can be seen in \eqref{eq:filter_coeff}, the intra-polarization coefficients are two times the inter-polarization coefficients, except for $n=0$ for SPM. This motivates the approximation of the relation between filter coefficients in \eqref{eq:filter_approximation}  in Section~\ref{s:linearchannel}.

\section*{Acknowledgments}
This work was supported by the Natural Sciences and Engineering Research Council of Canada (NSERC) and Huawei Tech., Canada. The research was enabled in part through support provided by the Digital Research Alliance of Canada (www.alliancecan.ca).

\bibliographystyle{IEEEtran}
\bibliography{references}

\begin{IEEEbiographynophoto}{Mohammad Taha Askari}
received the B.Sc. degree in electrical engineering from Sharif University of Technology, Tehran, Iran, in 2015, and the M.A.Sc. degree in electrical and computer engineering from the University of British Columbia, Vancouver, BC, Canada, in 2022. He is currently pursuing the Ph.D. degree in electrical and computer engineering at the University of British Columbia. His field of interest includes probabilistic shaping for optical fiber communications, machine learning for communications, and information and coding theory.
\end{IEEEbiographynophoto}

\begin{IEEEbiographynophoto}{Lutz Lampe}
received the Dipl.-Ing. and Dr.-Ing. degrees in electrical engineering from the University of Erlangen, Erlangen, Germany, in 1998 and 2002, respectively. Since 2003, he has been with the Department of Electrical and Computer Engineering, The University of British Columbia, Vancouver, BC, Canada, where he is a Full Professor. His research interests include theory and application of wireless, optical wireless, optical fiber, power line, and underwater acoustic communications. He has served as an associate editor and a guest editor for several IEEE journals, and as the general co-chair and the technical program committee co-chair for IEEE conferences. He has been a Distinguished Lecturer of the IEEE Communications Society and a (co-)recipient of a number of best paper awards. He is a Co-Editor of the book Power Line Communications: Principles, Standards and Applications from Multimedia to Smart Grid (2nd edition, John Wiley \& Sons).
\end{IEEEbiographynophoto}

\begin{IEEEbiographynophoto}{Jeebak Mitra}
currently works as a Member of Technical Staff in the Advanced Wireless Technologies (AWT) group at the office of the CTO, Dell Technologies, involved in design and development of next generation mobile network architectures and is responsible for various strategic initiatives related to Dell's wireless portfolio. Prior to that, Jeebak worked as a principal engineer at Huawei Technologies Research Center in Ottawa, Canada from 2013 to 2022, working as part of a world class DSP team that designed several generations of market leading optical DSP  transceivers focusing primarily on algorithm design and implementation for optical links for metro and transport network. From 2010 to 2013, he was a technical leader for DSP architecture and algorithms for wireless communications with several startup companies focused on wireless backhaul and access, all of which were acquired. He received the M.A.Sc. and Ph.D. degrees in Electrical Engineering from The University of British Columbia in 2005 and 2010, respectively. His research interests lie in the area of high-performance communication systems design focusing on wireless and optical transmission. Jeebak has over 20 issued patents and over 50 publications in various conferences and journals and received the Best Student Paper Award at the IEEE Canadian Conference in Electrical and Computer Engineering 2009. He was a co-recipient of the Best Paper Award at IEEE/ACM/IFIP CNSM 2017 and 2019.
\end{IEEEbiographynophoto}

\vfill

\end{document}